\documentclass[prx,aps,amssymb,nofootinbib,floatfix,reprint,notitlepage]{revtex4-2} 

\usepackage{times}
\usepackage[dvipsnames]{xcolor}
\usepackage{amsmath,bbold,bm}
\usepackage{mathtools}
\usepackage{graphicx}
\usepackage[normalem]{ulem}
\usepackage{cancel}
\usepackage{comment}
\usepackage{slashed}

\usepackage[pdfpagelabels,colorlinks=true,breaklinks=true,linktocpage=true,linkcolor=Blue,urlcolor=Blue,citecolor=Blue,pdfduplex=DuplexFlipLongEdge,%
pdftitle={Dynamic mass generation and topological order in overscreened Kondo lattices},%
pdfauthor={Yang Ge, Yashar Komijani}%
]{hyperref}

\newcommand{\be}{\begin{equation}}
\newcommand{\ben}{\begin{equation*}}
\newcommand{\ee}{\end{equation}}
\newcommand{\een}{\end{equation*}}
\newcommand{\bs}{\begin{split}}
\newcommand{\es}{\end{split}}
\newcommand{\bmx}{\begin{array}}
\newcommand{\emx}{\end{array}}

\newcommand{\bea}{\begin{eqnarray}}
\newcommand{\bean}{\begin{eqnarray*}}
\newcommand{\eea}{\end{eqnarray}}
\newcommand{\eean}{\end{eqnarray*}}
\newcommand{\dg}{^{\dagger}}
\newcommand{\dn}{^{\vphantom{\dagger}}}

\newcommand{\lr}{\leftrightarrow}

\newcommand{\bb}[1]{\mathbb{#1}}

\newcommand{\andd}{\qquad\text{and}\qquad}

\newcommand{\eps}{\epsilon}

\newcommand{\intoinf}[1]{\int_{0}^{\infty}{#1}}

\newcommand{\braket}[1]{\left\langle #1\right\rangle}

\newcommand{\mat}[1]{\left(\bmx{cc}#1\emx\right)}

\newcommand{\matn}[1]{\bmx{cc}#1\emx}

\newcommand{\bw}[1]{\begin{widetext}}
\newcommand{\ew}[1]{\end{widetext}}

\begin{document}

\title{Dynamic mass generation and topological order in overscreened Kondo lattices}
\author{Yang Ge}
\affiliation{Department of Physics, University of Cincinnati, Cincinnati, Ohio 45221, USA}
\author{Yashar Komijani$^{*}$}
\affiliation{Department of Physics, University of Cincinnati, Cincinnati, Ohio 45221, USA}
\date{\today}
\begin{abstract}
	Multichannel Kondo lattice models are examples of strongly correlated electronic systems that exhibit non-Fermi-liquid behavior due to the presence of a continuous channel symmetry. Mean-field analyses have predicted that these systems undergo channel symmetry breaking at low temperature. We use the dynamical large-$N$ technique to study temporal and spatial fluctuations of the multichannel Kondo model on a honeycomb lattice and find that this prediction is not generally true. Rather, we find a 2+1D conformally invariant fixed point, governed by critical exponents that are found numerically. When we break time-reversal symmetry by adding a Haldane mass to the conduction electrons, three phases, separated by continuous transitions, are discernible: one characterized by dynamic mass generation and spontaneous breaking of the channel symmetry, one where topological defects restore channel symmetry but preserve the gap, and one with a Kondo-coupled chiral spin liquid. We argue that the last phase is a fractional Chern insulator with anyonic excitations.
\end{abstract}

\maketitle

\section{Introduction}
The interplay between interaction and topology has emerged as a captivating field of study within condensed matter physics. Strong correlations can lead to new ground states which are endowed with exotic topological properties. Of particular significance is the realization of topological order \cite{Wen1990}, i.e.\ gapped insulating states with point-like excitations that have mutual nonabelian statistics. Such nonabelian anyons hold immense potential for applications in topological quantum computation \cite{Kitaev2003,Nayak2008}. Notable examples are nonabelian fractional quantum Hall states \cite{Moore1991,Read1999,Read2000}, quantum spin liquids \cite{Kitaev2006,Broholm2020} and topological superconductors \cite{Sato2017}. However, the unequivocal detection and precise control over these anyons have proven to be challenging. Consequently, the search for new platforms that can accommodate and manipulate topological order holds paramount importance, both from the fundamental understanding and practical application perspectives.

A corner stone of strongly correlated electronic systems is the Kondo effect, in which conduction electrons scatter off and screen a local magnetic moment. When several channels compete to screen the moment in the so-called multichannel Kondo model, a non-Fermi-liquid state arises \cite{Nozieres80} featuring fractional residual entropy at zero temperature, which is generally believed to arise from decoupled anyons \cite{Andrei84,Affleck92,Emery1992,Affleck1993}, and can potentially be used for topological quantum computation~\cite{Lopes20,Komijaniqubit,Rebecca21}.

\begin{figure}[th!]
	\includegraphics[width=1\linewidth]{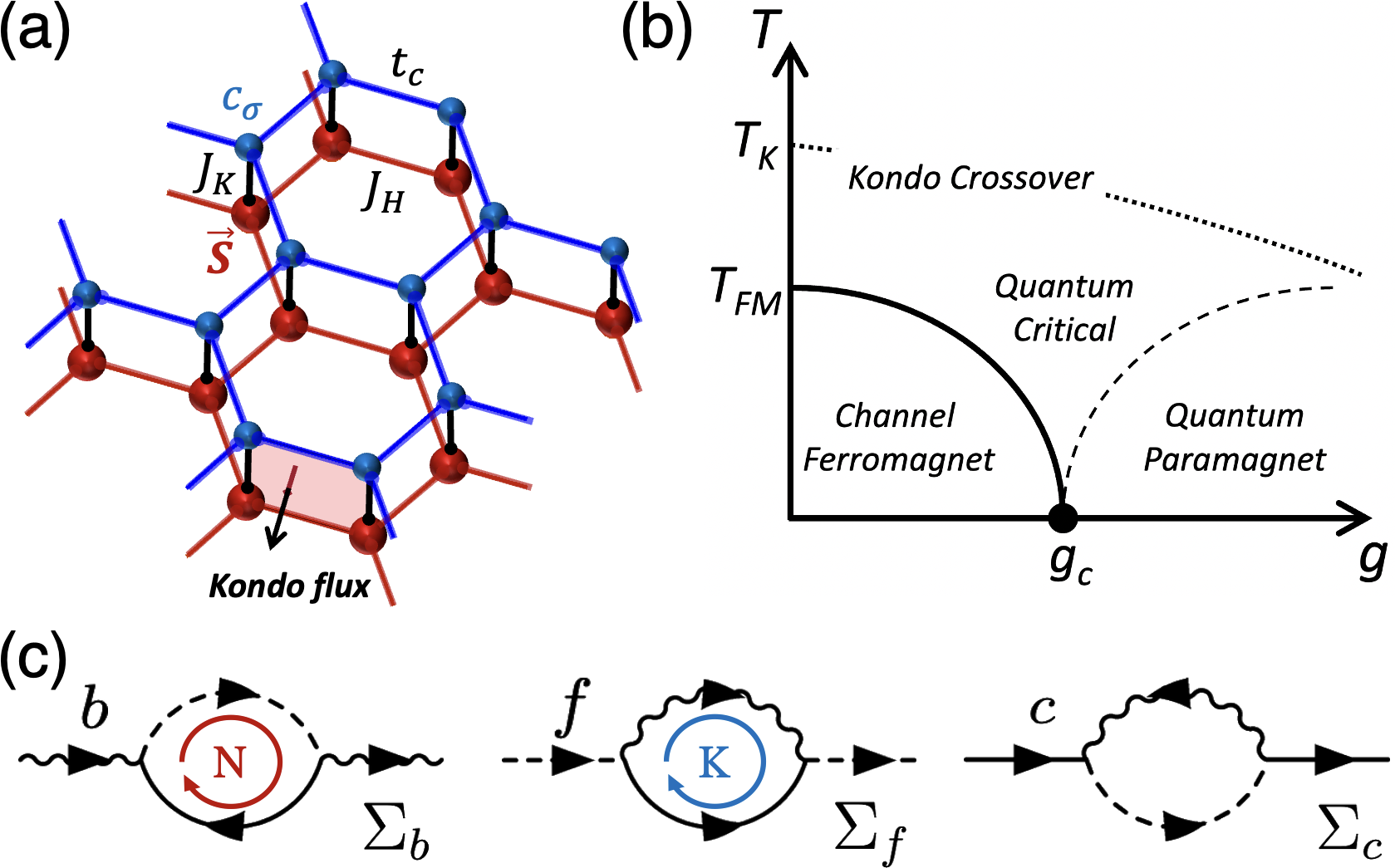}
	\caption{Kondo lattice and the two-channel phase diagram. (a) Sketch of the honeycomb Kondo lattice studied in this paper, which has two sublattice sites per unit cell. An example of Kondo flux, threading in-between conduction electrons and Abrikosov fermions (spinons) is illustrated. (b) The phase diagram previously proposed for two-channel Kondo lattice as a function of temperature $T$ and $g\sim T_K/J_H$\cite{Wugalter20}. (c) Self-energies from Kondo interaction for all the fields in the self-consistent equations. Each interaction vertex carries a $1/\sqrt{N}$ factor. While the internal sum over $\mathrm{SU}(N)$ spin or $\mathrm{SU}(K)$ channel indices brings $\Sigma_{b,f}$ to O(1), $\Sigma_c$ remains O$(1/N)$.
	}\label{fig1}
\end{figure}

While the case of a single impurity is well understood, much less is known about Kondo lattices where a lattice of spins is screened by conduction electrons. In the Kondo-dominated regime of a single-channel Kondo lattice, electrons form the familiar Fermi liquid but with a large Fermi surface (FS) \cite{Oshikawa00} and may even be driven to an insulating phase \cite{Hewson,Si2014,Coleman2015}. In the more complex case of a multichannel Kondo lattice, the continuous channel symmetry naturally leads to new patterns of entanglement which are potentially responsible for the non-Fermi-liquid physics \cite{Jarrell96,Jarrell1997}, symmetry breaking, and fractionalized order parameters \cite{ofc}.

The channel symmetry breaking in a multichannel Kondo lattice (MCKL) \cite{Cox1996} has been the focus of a number of recent studies. When the symmetry breaks spontaneously, the system becomes effectively a single-channel Kondo lattice since all but one channels decouple entirely. %
Such symmetry broken states are relevant to experiments and real materials, as the MCKL, and in particular the two-channel Kondo lattice (2CKL), seem to be appropriate models for several heavy-fermion compounds, e.g., the family of Pr\textsl{Tr}$_2$Zn$_{20}$ (\textsl{Tr}=Ir,Rh)~\cite{Onimaru2019,Patri2020}. Moreover, due to the natural frustration of the channel degree of freedom, one can speculate that certain deformations of the MCKL may realize a topological order \cite{Wen1990} with anyonic excitations. One of the simplest deformations is inducing a topology via the conduction electrons. Motivated to explore exotic phases as a result of such deformation, we plunge into a detailed study of the phase diagram of the multichannel Kondo lattice model on a honeycomb lattice with and without time-reversal symmetry.

The MCKL model is described by the Hamiltonian 
\begin{equation}
H=H_c+ J_H \sum_{\langle ij \rangle}\vec S_{i}\cdot\vec S_{{j}}+J_K\sum_{j}\vec S_{j}\cdot c\dg_{{j}a\alpha}\vec\sigma\dn_{\alpha\beta} c\dn_{{j}a\beta},
\label{eqH}
\end{equation}
where $H_c=\sum_{ij}(-t^{ij}_c c\dg_{{i}a\alpha}c\dn_{{j}a\alpha}+\mathrm{h.c.})$ is the Hamiltonian of the conduction electrons, the $J_H$ term describes antiferromagnetic Heisenberg interaction between the nearest-neighbor (NN) spins, the $J_K$ term describes the Kondo interaction, and the Einstein summation convention over spin $\alpha,\beta=1,\dots,N$ and channel $a,b=1,\dots,K$ indices is assumed. This is schematically represented in Fig.\,\ref{fig1}(a). The model has $\mathrm{SU}(N)$ spin and $\mathrm{SU}(K)$ channel symmetries, and we are interested in the possible spontaneous channel symmetry breaking with an order parameter $\vec {\cal O}_{ j}\equiv (\vec S_{ j}\cdot c\dg_{ ja\alpha}\vec\sigma\dn_{\alpha\beta} c\dn_{ jb\beta}) \vec \tau\dn_{ab}$ where $\vec\tau$-s act as $\mathrm{SU}(K)$ generators in the channel space \cite{Hoshino11}.

A direct solution of the model \eqref{eqH} is in general impossible due to the strongly correlated nature of the problem, and 
previous attempts have led to conflicting results. In a strong Kondo coupling limit the model is described by the effective Hamiltonian $H_\text{eff}=\lambda\sum_{\braket{ij}}\vec{\cal O}_i\cdot\vec{\cal O}_j$, favoring a channel antiferromagnet when $\lambda>0$ \cite{Schauerte05,Ge22}.
On the other hand, mean-field studies based on the large-$N$ limit predict a variety of channel ferromagnetic and channel antiferromagnetic solutions depending on the conduction filling \cite{vanDyke19,Zhang18,Wugalter20}.  Early single-site dynamical mean-field theory (DMFT) studies predicted non-Fermi-liquid physics but could not go to low temperatures \cite{Jarrell96,Jarrell1997}. However, more recent studies provided evidence for a channel symmetry broken phase \cite{Hoshino11,Kuramoto}. Interestingly, this has {not been confirmed} in recent cluster DMFT studies~\cite{Inui20}. They found that as the polarizing field goes to zero, the order parameter goes to zero as well. Clearly, spatial fluctuations are playing a role in destabilizing the ordered phase. 

The effective theory of fluctuations in 2CKL was studied in Ref.\,\onlinecite{Wugalter20} in the large-$N$ limit. It can be formulated as ${\cal L}={\cal L}_{\text{NL}\sigma\text{M}}+{\cal L}_\text{fermions}$ where
\begin{equation}
{\cal L}_{\text{NL}\sigma\text{M}}=\frac{(\partial_\mu \vec n)^2}{g}+\Gamma(A^{\text{ex}}_\mu-A^{\text{in}}_\mu-\Omega^z_\mu)^2.\label{eq2}
\end{equation}
The first term is the usual nonlinear sigma model (NL$\sigma$M) governing the fluctuations of the order parameter $\vec n\sim\vec{\cal O}$. 
The second term in \eqref{eq2} is a Higgs term with stiffness $\Gamma$ that tends to lock the internal U(1) gauge field of $f$-electrons $A^{\text{in}}_\mu$ with the external electromagnetic gauge potential $A^{\text{ex}}_\mu$ up to ${\Omega}^z_\mu$.
At low temperatures, the NL$\sigma$M has an ordered phase where channel symmetry is spontaneously broken [Fig.\,\ref{fig1}(b)], and a quantum paramagnet phase in which the symmetry is restored by topological defects (channel skyrmions). Here $\Omega^z_\mu$ is the gauge-dependent disorder potential, whose curl gives the density of defects $4\eps^{\vphantom{z}}_{\mu\nu\lambda}\partial^{\vphantom{z}}_\nu \Omega_\lambda^z=\eps_{\mu\nu\lambda}\vec n\cdot(\partial_\nu\vec n\times\partial_\lambda\vec n)$. 
In the ordered phase, defects are expelled and the internal and external gauge fields are locked leading to a large FS, whereas in the quantum paramagnet phase the defects proliferate destroying the coherent phase-locking \cite{Wugalter20,Grover2008}.

However, since the winning channel has a larger FS \cite{ofc,Wugalter20} and the order parameter $\vec{\cal O}$ is strongly dissipated by coupling to fermionic degrees of freedom ${\cal L}_\text{fermions}$, the ground state is expected to be more complicated, at least in two or three spatial dimensions. This is reminiscent of the spin-fermion model where the gapless fermionic modes need to be explicitly treated in the renormalization group (RG) study \cite{Millis92,Altshuler95,Sachdev95,Abanov2003,Oganesyan01,Metzner03,Chowdhury14,Mross10,Metlitski15,Mahajan13,Fitzpatrick13,Fitzpatrick14,Torroba14,Fitzpatrick15,Metlitski10,Lee09,Dalidovich13,Lee2018}. Thus, an unbiased and more systematic approach is needed to uncover the physics of MCKL.
 
Encouraged by recent successes of the large-$N$ approach to Kondo lattices as cross-checked by tensor network \cite{Chen2023} and Quantum Monte Carlo \cite{Raczkowski2022} methods, we conduct a large-$N$ study of MCKL.
In a previous study \cite{Ge22}, we applied the dynamical large-$N$ approach to both 1D and $\infty$D MCKLs with Schwinger boson representation of spins and showed that spatial fluctuations are fully captured by including momentum dependence of the self-energy. 

In this paper, we use this state-of-the-art technique to shed light on the issue of the symmetry breaking in the MCKL in 2+1D, which lies between upper and lower critical dimensions. By focusing on a honeycomb lattice we show that the channel symmetry is unbroken and the ground state is a channel symmetry-preserving 2+1D conformal fixed point. We then probe the fate of this state under the breaking of time-reversal symmetry which induces a topological gap. Remarkably this leads to three different states, two of which resemble the phases in the NL$\sigma$M picture above, i.e.\ a spontaneous channel symmetry broken state and a gapless quantum paramagnet. In addition, we show that there is a third gapped phase with fractional edge states that has topological order.

While the formalism used is that of Ref.~\onlinecite{Ge22}, there are several differences. In this paper we use the Abrikosov fermion representation of the spin instead of the Schwinger boson representation. The previous paper discussed 1+1D and $\infty$+1D cases, whereas the current paper focuses on 2+1D. Finally, Ref.~\onlinecite{Ge22} was more concerned about breaking the continuous channel symmetry, whereas here we break time-reversal symmetry using the Haldane fluxes in the honeycomb lattice, and (in one of phases) the channel symmetry breaks spontaneously as a result.

This paper is organized as follows: In Section \ref{ss:method}, we introduce the dynamical large-$N$ formalism utilized throughout this study. Section \ref{ss:res} details our numerical findings for both time-reversal symmetric gapless and time-reversal symmetry broken gapped conduction electrons. An interpretation of these numerical results, along with the phase diagram of the system, is presented in Section \ref{ss:interp}. The paper concludes in Section \ref{ss:discussion}, where we offer additional remarks and highlight some unresolved questions. The appendixes supplement the main text with detailed information. Appendix \ref{ss:app:1d} provides an overview of the 1+1D results related to \cite{Ge22}, while Appendix \ref{ss:app:numerics} delves into the specifics of our numerical simulations. Appendix \ref{ss:app:cft} explores our conformal scaling ansatz in 2+1D and its application to our results. Finally, Appendix \ref{ss:app:gapped} further examines the various phases encountered in scenarios with gapped conduction electrons.

\begin{figure}[tp!]
	\includegraphics[width=1\linewidth]{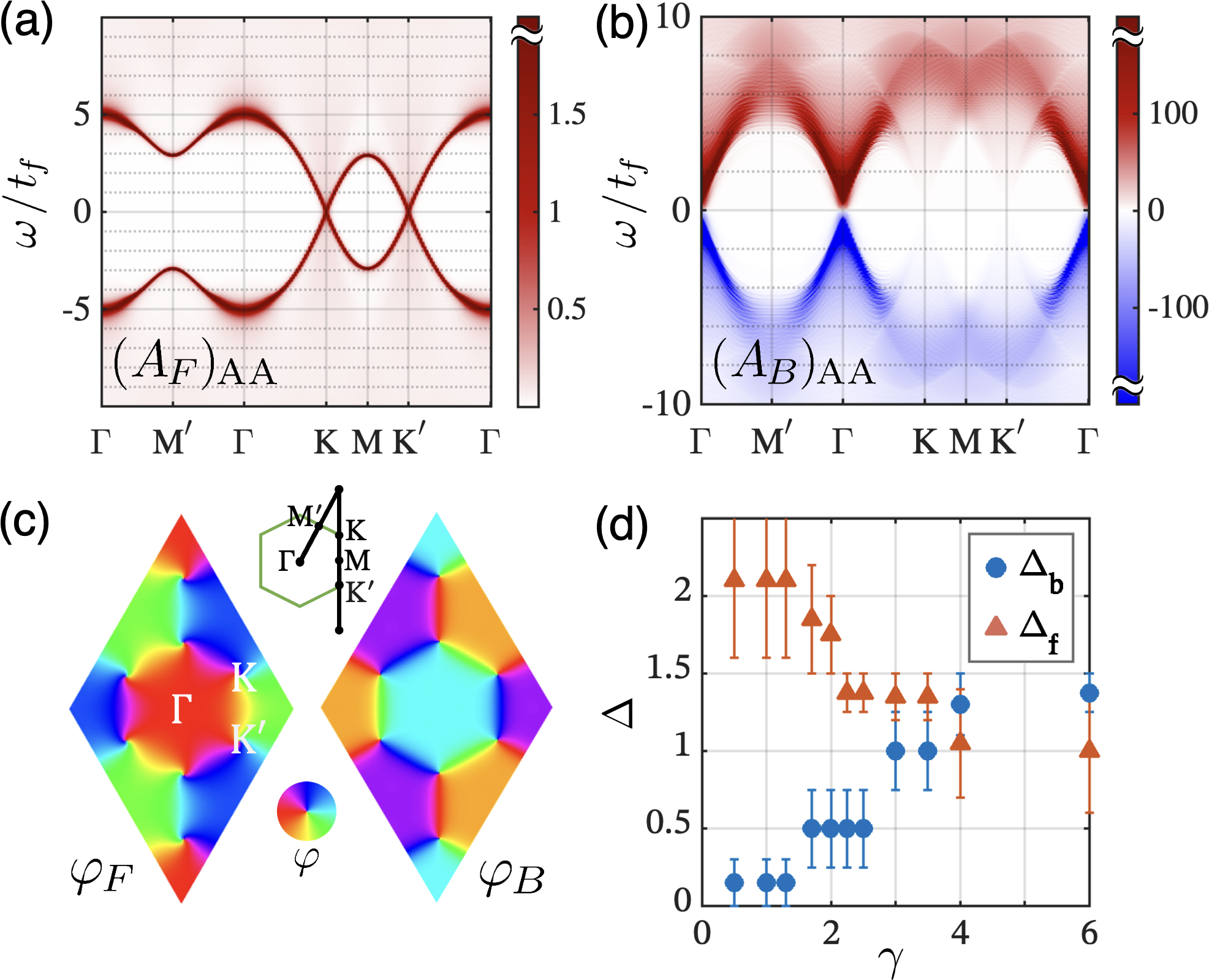}
	\caption{Low energy conformal modes of spinons and holons with gapless conduction channels. The spectral functions of (a) spinons, $A_F$ and (b) holons, $A_B$, exhibit lightcones with the same group velocity at low temperature ($T/J_K=0.005$) along the Brillouin-zone cut shown below (a). (c) The phase structure in the critical eigenstates of $G_f(0+i\eta,k)$ and $G_b(0+i\eta,k)$, extracted from the angle $\varphi$ between the $\sigma^{x,y}$ components in a $C_3$-symmetric reciprocal-space gauge. The vortices and plains show a pair of critical chiral fermions at $\mathrm{K}$ and $\mathrm{K}'$ as well as a chiral boson at $\Gamma$. Parameters used in (a--c) are $\gamma=4$, $J_K/t_c=10$, and $t_f/t_c=-0.2$. (d) Critical exponents of spinons and holons extracted from scaling analysis.}\label{fig2}
\end{figure}

\section{Dynamical large-$N$ method for the multichannel Kondo lattice \label{ss:method}}%

We assume that the spins transform according to a fully antisymmetric representation of $\mathrm{SU}(N)$ given by the Abrikosov fermions \cite{Abrikosov1965,Affleck1985,Affleck1988,Arovas1988,Coleman2015}, so that $S_{j\alpha\beta}=f\dg_{j\alpha}f\dn_{j\beta}$, with the constraint $f\dg_{ j\alpha}f\dn_{ j\alpha} =N/2$, forming a self-conjugate representation of spins. This introduces a gauge symmetry $f_{ j\alpha}\to e^{i\varphi_{ j}}f_{ j\alpha}$, which we track throughout the paper. We maintain the particle-hole (p-h) symmetry for both spins and conduction electrons, which implies that the constraint is satisfied on average.  In the impurity case \cite{Jerez1998,Parcollet1998}, the spin is overscreened for any number of channels when $K>1$. We rescale $J_K\to J_K/N$ and send $N,K\to\infty$ while keeping the ratio $\gamma=K/N$ constant.  After a Hubbard-Stratonovich (HS) decoupling of the Heisenberg and Kondo interaction, the Lagrangian becomes ~\cite{Parcollet1997,Komijani18}
\begin{eqnarray}
	\label{eq3}
	L &=& L_c+L_f+L_K,\nonumber\\
	L_c&=& \sum_k \sum_{mn} \bar{c}^m_{ka\alpha} [\partial_\tau {\bb 1} + H_c(k)]_{mn} c^n_{ka\alpha}, \nonumber\\
	L_f&=& \sum_i \bar{f}_{i\alpha} \partial_\tau f_{i\alpha} + \sum_{\langle ij \rangle} \bigg[ \frac{N\vert{t_f^{ij}}\vert^2}{J_H}-(t_f^{ij}\bar{f}_{i\alpha} f_{j\alpha}+\mathrm{h.c.})\bigg],  \nonumber\\
	L_K&=&\sum_{i}\bigg[\frac{\bar b_{ia}b_{ia}}{J_K}+\frac{1}{\sqrt N}(\bar c_{ia\alpha}f_{i\alpha}\bar b_{ia}+\mathrm{h.c.})\bigg].
\end{eqnarray}
Here, $b_{ia}$ are bosonic holon fields introduced for the HS decoupling of the Kondo interaction, and the NN spinon hopping $t^{ij}_f$ comes from the HS decoupling of the Heisenberg interaction. The bare Hamiltonian $H_c(k)$ governs the conduction electrons, which can include sublattice indices $m$ or $n$, and $\mathbb{1}$ denotes the identity matrix in the sublattice space. We further assume that $t^{ij}_f=t_f$ is uniform \cite{Arovas1988}.

The gauge symmetry $f_{ j}\to e^{i\varphi_{ j}}f_{ j}$ is preserved in \eqref{eq3}, provided that $b_j\to e^{i\varphi_{ j}}b_{ j}$. The channel order parameter $\vec{\cal O}_{ j}$ in terms of the holon fields becomes $\vec{\cal B}\dn_{ j}\equiv b\dg_{j a}\vec\tau\dn_{a b}b\dn_{ j b}$, as we showed before \cite{Ge22}.

In the overscreened case, any infinitesimal $J_H$ delocalizes spinons due to a resonant Ruderman-Kittel-Kasuya-Yosida (RKKY) amplification \cite{Ge22} and produces a finite $t_f$, (see Appendix \ref{ss:app:scaling}). In the rest of this paper, $J_H$ is typically chosen to be small enough so that this emergent contribution is the dominant source of spinon delocalization at low temperature.

In the large-$N$ limit, this system \eqref{eq3} can be solved exactly as the dynamics is dominated by noncrossing Feynman diagrams [Fig.\,\ref{fig1}(c)], resulting in the imaginary-time self-energies
\begin{eqnarray}
\label{eqself}
\Sigma_f(\mathbf{r}_i,\tau)&=&-\gamma g_c(\mathbf{r}_i,\tau)G_b(\mathbf{r}_i,\tau),\nonumber\\
\Sigma_b(\mathbf{r}_i,\tau)&=&g_c(\mathbf{r}_i,\tau)G_f(\mathbf{r}_i,\tau),
\end{eqnarray}
whereas $\Sigma_c \sim \mathrm{O}(1/N)$. Hence, the electron propagator remains bare.
In terms of crystal momentum and the complex frequency $\mathrm{z}$, and in the sublattice basis, the Dyson equations read
\begin{eqnarray}
\label{eqdyson}
	g_c^{-1}(\mathbf{k},{\rm z})&=&{\rm z}\mathbb{1}-{H}_c(\mathbf{k}), \nonumber\\
	G_f^{-1}(\mathbf{k},{\rm z})&=&{\rm z}\mathbb{1}-H_f(\mathbf{k})-\Sigma_f(\mathbf{k},{\rm z}),\nonumber\\
	G_{b}^{-1}(\mathbf{k},{\rm z})&=&-J_{K}^{-1}\mathbb{1}-\Sigma_b(\mathbf{k},{\rm z}),
\end{eqnarray}
where $H_f$ is the spinon-hopping Hamilton due to the uniform $t_f$.
Together with Eq.\,\eqref{eqself}, they form a set of coupled integral equations that are solved iteratively and self-consistently to extract the thermodynamics \cite{Rech2006,Komijani18}. To tame numerical complexity, symmetries are utilized to reduce the computation down to the fundamental domain of the Brillouin zone (see Appendix \ref{ss:app:numsym}) \cite{Monkhorst1976,kruthoff_topological_2017}.

\section{Results \label{ss:res}}

In this study we focus on the 2+1D MCKL on a honeycomb lattice. The 1+1D 2CKL was previously studied in Ref.\,\onlinecite{Ge22} using a dual representation, whose ground state is a conformally invariant fixed point. In addition to scale and translational invariances, such fixed points exhibit emergent Lorentz invariance and are additionally invariant under the special conformal transformations. The spinons and holons have critical exponents $\Delta_b=\frac{\gamma}{\gamma+2}$ and $\Delta_f=\frac{1}{2}+\frac{2}{\gamma+2}$, as detailed in Appendix \ref{ss:app:1d}.

On the honeycomb lattice, we consider the Haldane model of conduction electrons \cite{Haldane1988}. In the $\mathrm{A}$--$\mathrm{B}$ sublattice basis,
\begin{equation}
\label{eq:Hc}
H_c (\mathbf{k})\!=\!\begin{pmatrix}
	- 2| t_c' | \sum_{j=1}^3 \sin (\mathbf{k} \cdot \mathbf{s}_j)  & - t_c e^{i \sum_{j=1}^3 \mathbf{k} \cdot \mathbf{d}_j } \\
	\mathrm{c.c.} & 2| t_c' | \sum_{j=1}^3 \sin (\mathbf{k} \cdot \mathbf{s}_j)
\end{pmatrix}\!,
\end{equation}
where $\mathbf{d}_j, j=1,2,3$ are the displacements to the three NN $\mathrm{B}$ sites of $\mathrm{A}$, and $\mathbf{s}_j=\epsilon_{jlm}(\mathbf{d}_l-\mathbf{d}_m)/2$ are three of the displacements to next-nearest-neighbor (NNN) sites. Here $t_c'$'s are the purely imaginary NNN hoppings that generate the Haldane mass. In general the honeycomb lattice also admits a Semenoff mass staggering between the A and B sublattice sites of the honeycomb, i.e., $m^S_c \big( \sum_{i\in \mathrm{A}}c\dg_i c\dn_i-\sum_{i\in \mathrm{B}}c\dg_i c\dn_i \big)$.
We exclude the Semenoff mass to enforce p-h symmetry on every site.

We define the sublattice spinors of the fields as
\begin{equation}
	\renewcommand*{\arraystretch}{1.2}
	C_{a\alpha}=\begin{pmatrix}
		c_{a\alpha}^\mathrm{A} \\
		c_{a\alpha}^\mathrm{B}
	\end{pmatrix},
	\hspace{0.25cm}
	F_{\alpha}=\begin{pmatrix} f_{\alpha}^\mathrm{A} \\ f_{\alpha}^\mathrm{B}\end{pmatrix},
	\hspace{0.25cm}
	B_{a}=\begin{pmatrix} b_{a}^\mathrm{A} \\ b_{a}^\mathrm{B}\end{pmatrix},\label{eq:spinor}
\end{equation}
where the superscripts $\mathrm{A},\mathrm{B}$ denote the corresponding sublattice sites that the fields reside on. Henceforth, we denote by $\vec\sigma\equiv(\sigma^x,\sigma^y,\sigma^z)$ the Pauli matrices that act in the sublattice space. The conjugates of the spinors above are defined by 
\begin{equation}
	\bar{C}\dn_{a\alpha} = C\dg_{a\alpha} \sigma^z, \quad \bar{F}\dn_{\alpha} = F\dg_{\alpha} \sigma^z, \quad \bar{B}\dn_a = B\dg_a.
\end{equation}
Furthermore, we denote their low-energy modes at low temperature by 
\begin{equation}
	\label{eq:lowE}
	C_{a\alpha}\sim\psi_{a\alpha}, \quad F_{\alpha} \sim \chi_{\alpha}, \quad B_a \sim \phi_a.
\end{equation}
From now on, spin and channel indices will be suppressed when they are inessential.

\subsection{Gapless conduction channels \label{ss:gapless}}

First, we assume that conduction electrons have only NN hoppings as in graphene \cite{Neto2009}, i.e., $t_c'=0$ in $H_c$ \eqref{eq:Hc}. The low energy $c$-electrons consist of two chiral spinors each sitting at the two Dirac nodes, $\psi_{\rho}$ for $\rho=1,2$, located at crystal momenta $\mathrm{K}$ and $\mathrm{K}'$,
\begin{equation}
C(\mathbf{r},\tau)=e^{-i\mathbf{K}\cdot \mathbf{r}}\psi_1(\vec{r}\,)+e^{-i\mathbf{K}'\cdot\mathbf{r}}\psi_2(\vec{r}\,),
\end{equation}
where $\vec r \equiv (x,y,v\tau)\equiv(\mathbf{r},v\tau)$. Furthermore, define  $\vec\nabla=(\partial_x,\partial_y,v^{-1}\partial_\tau)$, $\cancel r = \vec\sigma\cdot\vec r$ and $\cancel\partial=\vec\sigma\cdot\vec\nabla$ for one Dirac cone, and $\bcancel r = \vec\sigma^*\cdot\vec r$ and $\bcancel\partial=\vec\sigma^*\cdot\vec\nabla$ for the other which has opposite chirality. 

Numerical results indicate the presence of an intermediate fixed point where $f$-electrons develop Dirac dispersion at $\mathrm{K}$ and $\mathrm{K}'$. In this regime, only the energy scale close to the $c$- and $f$-Dirac nodes are relevant. Although the numerical calculations are all performed on the original tight-binding model \eqref{eq3}, for future references we note that near this intermediate fixed point, the Euclidean Lagrangian density $\mathcal{L}'=\mathcal{L}'_0+\mathcal{L}'_\text{int}$ can be expressed in terms of the low-energy modes \eqref{eq:lowE},
\begin{eqnarray}
\label{eq:L*}
{\cal L}'_{0} &=& \bar\psi_1\cancel{\partial}_c\psi_1+\bar{\psi}_2\bcancel{\partial}_c\psi_2+\bar \chi_1\cancel{\partial}_f \chi_1+\bar{\chi}_2\bcancel{\partial}_f\chi_2+m_b\bar\phi\phi, \nonumber\\
{\cal L}'_\text{int} &=& \frac{1}{\sqrt N} \left(\psi^{\dagger}_{1a\alpha} \phi^{\dagger}_a \chi\dn_{1\alpha} + \psi^{\dagger}_{2a\alpha} \phi^{\dagger}_a \chi\dn_{2\alpha}\right)+\mathrm{h.c.}
\end{eqnarray}
Here $\chi_{1,2}$ are the spinon counterparts to $\psi_{1,2}$, also at $\mathrm{K}$ and $\mathrm{K}'$ respectively. On the other hand, the only low-energy mode of the holon $\phi$ is at the $\Gamma$ point. The derivatives $\cancel\partial_{c}$ and $\cancel\partial_{f}$ correspond to $\cancel\partial$ with group velocities $v_{c}$ and $v_{f}$ respectively. In addition to the SU($N$) spin, SU($K$) channel, internal U(1) gauge, and p-h symmetry, the action has time-reversal (TR) symmetry $i\to -i$  and inversion (I) symmetry $(x,y)\to-(x,y)$, which act on spinors according to
\begin{equation}
{\rm TR:}\quad \matn{\psi_1\lr\psi_2\\ \chi_1\lr \chi_2}, \andd {\rm I:}\quad\matn{\psi_1\lr\sigma^x\psi_2 \\ \chi_1\lr \sigma^x\chi_2}.
\end{equation}
Both symmetries leave $\phi$ unchanged.
The electron Green's function is
\begin{equation}
g_C(\vec{r}\,)=\frac{1}{4\pi r^{{3}}}\Big(e^{-i\mathbf{K}\cdot \mathbf{r}} \, {\cancel r}+e^{-i\mathbf{K}'\cdot \mathbf{r}}{\bcancel r}\Big). \label{eq:gc}
\end{equation}
Figures \ref{fig2}(a)(b) show the spectral functions of spinons and holons, respectively. At a first glance, the spectral function of Abrikosov fermions resembles that of a noninteracting fermion: gapless at $\mathrm{K}$ and $\mathrm{K}'$ with opposite phase windings shown in Fig.\,\ref{fig2}(c). Furthermore, its bandwidth is resonantly amplified, similar to the case of Schwinger bosons~\cite{Ge22}. 

However, there is a crucial difference, namely the strong incoherent contribution inside the light cones. In fact, this interaction-driven fixed point exhibits a 2+1D criticality, which leads to various power-law spectra at low frequencies.  We find that Green's functions are in good agreement with the conformally invariant ansatzes
\begin{eqnarray}
G_F(\vec r\,)&=&\frac{\alpha_f}{4\pi r^{{2\Delta_f+1}}}\Big(e^{-i\mathbf{K}\cdot \mathbf{r}}{\cancel r}+e^{-i\mathbf{K}'\cdot \mathbf{r}}{\bcancel r}\Big) \label{eq:Gf}\nonumber \\
G_B(\vec r\,)&=&\frac{\alpha_b}{4\pi r^{{2\Delta_b}}}{\cal P}_b\label{eq:Gb},
\end{eqnarray}
where $\alpha_{b,f}$ are constants, and ${\cal P}_b=\frac{1}{2}(\mathbb{1}+\sigma^x)$ is a projector to the symmetric bonding state between the two sublattice sites.  This projection is reflected in the observation that only uniform (rather than sublattice-staggered) channel susceptibility \cite{Ge22} can diverge at low temperature. Therefore, the low energy spinor for the holon is $\phi \equiv \mathcal{P}_b B$.

It is important to note that the conformal invariance of Eqs.\,\eqref{eq:Gb}, including their Lorentz symmetry, was absent at the UV and is an emergent property of the IR fixed point. We should also note that such a fixed point cannot be accessed via local approximations, like DMFT or any of its extensions.

Our numerical scaling analysis (Appendix \ref{ss:app:scaling}) gives the critical exponents $\Delta_f$ and $\Delta_b$ as shown in Fig.\,\ref{fig2}(d). In particular, $\lim_{\gamma\to0}\Delta_b=0$, corresponding to a constant hybridization in the perfect screening case at $K=1$ (see also Appendix \ref{ss:app:1dCFT}). These exponents are governed by the conformally invariant fixed point.

To summarize, the ground state of the model is governed by a conformally invariant fixed point ${\cal L}_*$ that preserves the channel symmetry. This is true for any $J_K$, in marked contrast to large-$N$ static mean-field theory and single-site DMFT results.  In other words, we find that in the presence of TR symmetry our large-$N,K$ limit predicts a SU$_\text{ch}$(K) symmetry-preserving ground state irrespective of the value of $J_K$ and $K/N$. In the finite-$K$ studies \cite{Zhang18,Wugalter20,Kuramoto}, the bosons (which carry the channel quantum number) are condensed from the onset. In the NL$\sigma$M study of Ref.\,\onlinecite{Wugalter20} the fluctuations in the magnitude of the bosonic spinorial order parameter are gapped, whereas its direction undergoes a spontaneous symmetry breaking at low temperature. In our study the bosons evade condensation by remaining critical down to $T=0$. We attribute  criticality to the coupling of the channel order parameter to the gapless conduction electrons and speculate that this result applies to other large-$K$ MCKLs. In such systems symmetry breaking can be induced by asymmetric couplings of the channels \cite{Ge22}, but does not occur spontaneously, in agreement with cluster DMFT results \cite{Inui20}. Whether or not this large-$K$ prediction of symmetry preservation holds for a finite $K$ number of channels, is left for future studies.

\begin{figure}[tp!]
	\includegraphics[width=1\linewidth]{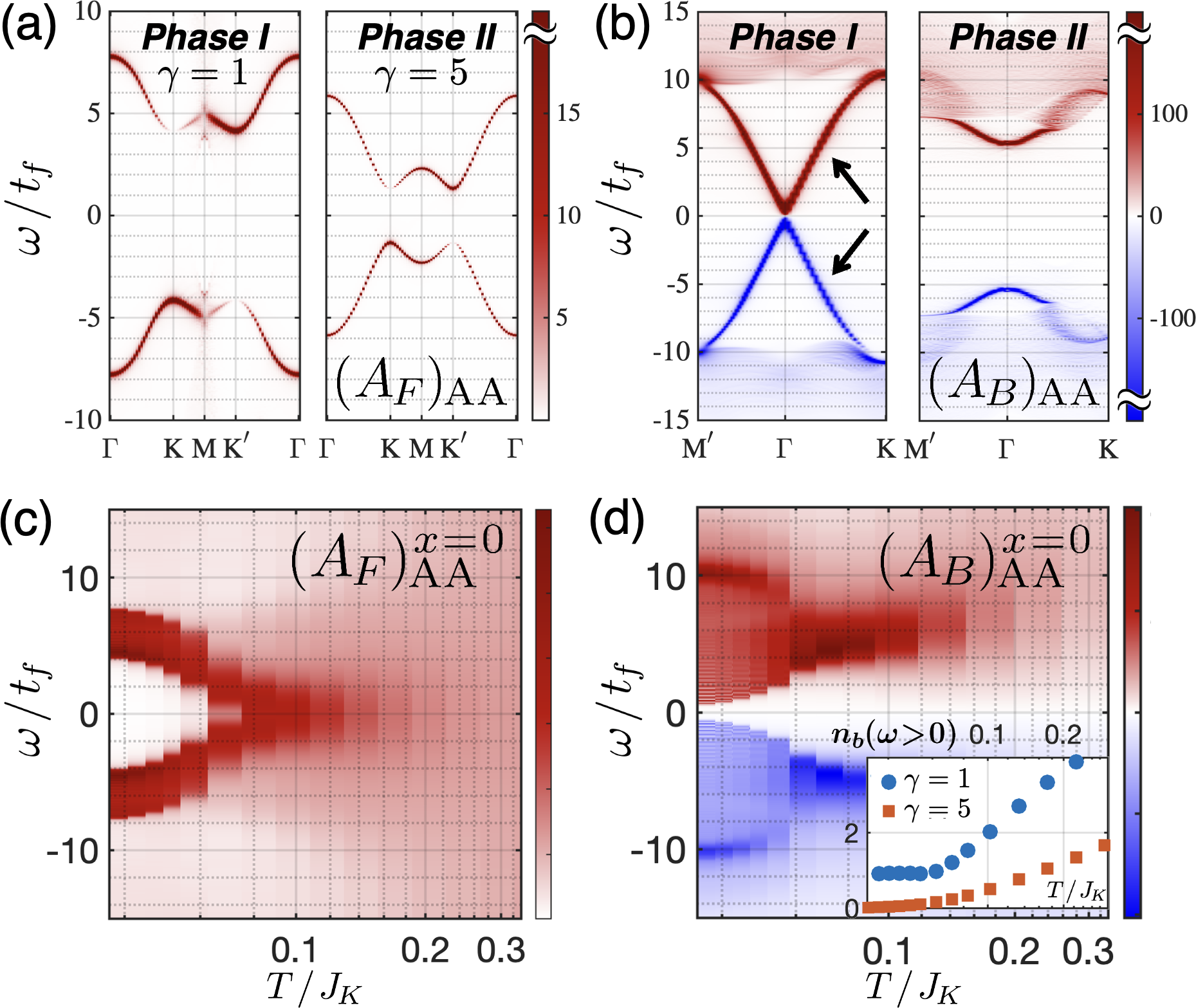}
	\caption{\label{fig3}
		Spectral functions of (a) spinons, $A_F$ and (b) holons, $A_B$ coupled to Haldane mass-gapped conduction channels. In Phase I ($\gamma=1$) the holon gap is vanishing, while in Phase II ($\gamma=5$) both spinons and holons are gapped. The arrows in (b) point to a true p-h symmetric bound state $\phi \sim \psi\dg \chi$ in an otherwise gapped spectrum. Spectra in (a) and (b) are taken at $T/J_K=0.03$. (c--d) Temperature evolution of the local spectral functions in Phase I, of (c) spinons and (d) holons. Inset of (d) shows the populations of free holons in the two cases, which for $\gamma=1$ becomes conserved at low temperature. Parameters used are $J_K/t_c=6$, $|t_c^{\prime}/t_c|=0.5$, and $t_f/t_c=-0.2$.
		}
\end{figure}

\subsection{Gapped conduction channels with Haldane mass \label{ss:gapped}}

Next, we break the TR symmetry by adding purely imaginary NNN hoppings \emph{only} in the conduction-electron layer, setting $t_c'\neq0$ in $H_c$ in Eq.\,\eqref{eq:Hc}. This turns the $c$-layer into a Haldane model, where a periodic magnetic flux goes through the honeycomb yet the flux totals to zero in each unit cell \cite{Haldane1988}.
This gaps out the conduction electrons with the Haldane mass. Remarkably, the $f$-spinons inherit a similar gap via the phenomenon of resonantly enhanced dispersion we reported before \cite{Ge22} [Fig.\,\ref{fig2}(a)]. The form and the nature of the gap, however, depends on the parameter $\gamma=K/N$ and various distinct phases are discernible, as shown in Figs.\,\ref{fig3}(a) and \ref{fig3}(b).
At large $\gamma>2$, labeled Phase II, we find that fermionic spinons and bosonic holons are both gapped. Their spectral gap remain independent of temperature at low $T$.
At smaller $\gamma<2$, labeled Phase I, spinons have a much different shape of dispersion compared with Phase II while holons have a spectral gap that depends on $T$.  Their spectral evolution with $T$ in Phase~I are shown in Figs.\,\ref{fig3}(c) and \ref{fig3}(d). As $T$ decreases, a gap opens in the local spectrum of spinons. On the other hand, the gap in the holon spectrum shrinks with reducing $T$, so that as $T\to0$ the ground state becomes gapless.

\begin{figure}[tp!]
	\includegraphics[width=1\linewidth]{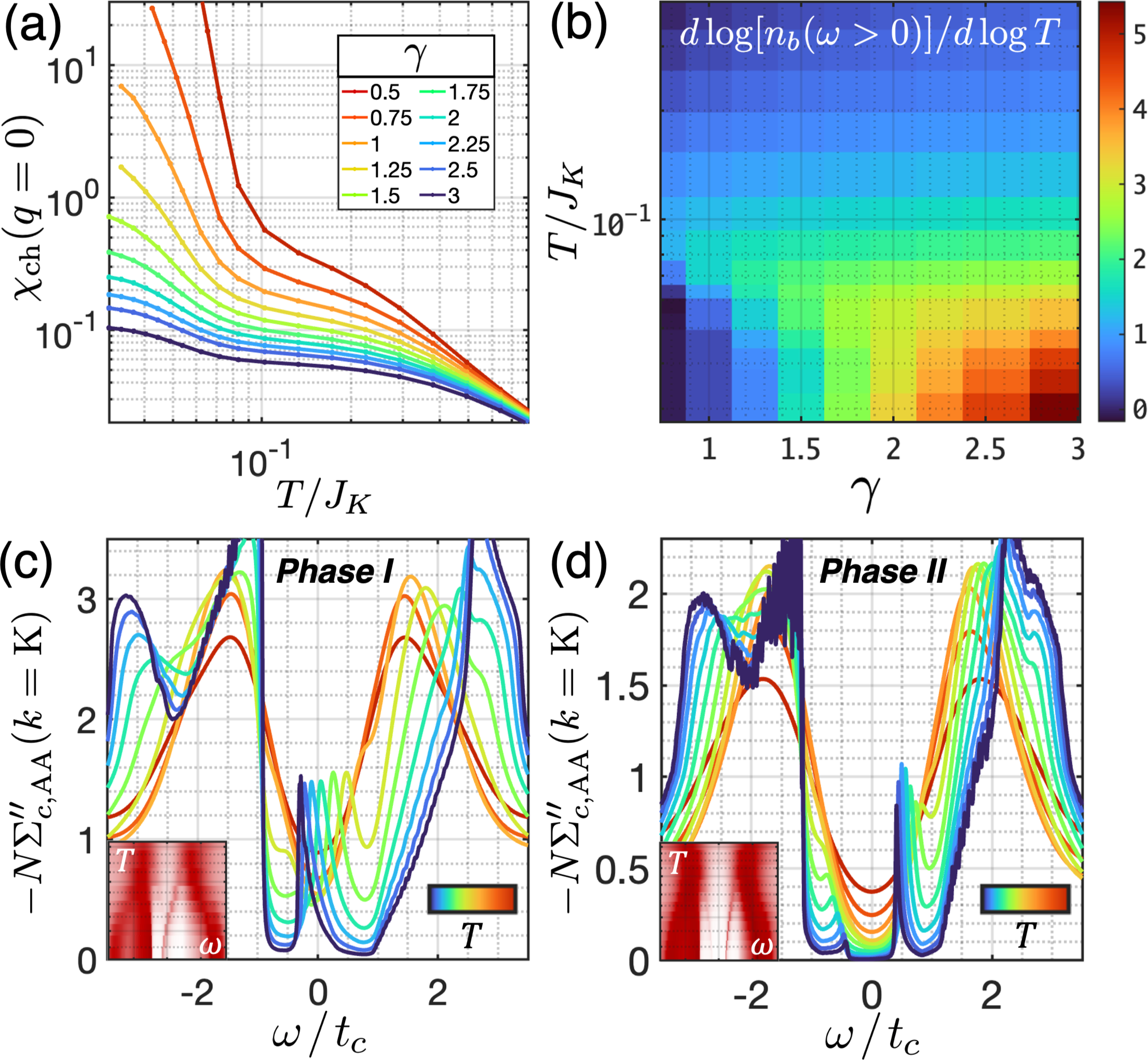}
		\caption{\label{fig4}
		Channel susceptibility, free holon population and the conduction-electron self-energy across the phase diagram. 
		(a) Uniform channel susceptibility $\chi_{\text{ch}}(q\!=\!0)$ vs $T$ at several $\gamma$, at a fixed $t_f=-0.4t_c$, which fixes $J_H$ at low $T$.  Low-temperature $\chi_{\text{ch}}$ diverges in Phase Ia, signaling spontaneous channel symmetry breaking and the formation of a channel ferromagnet.
		(b) Temperature dependence of the free holon population, represented by $\xi(T)\equiv \mathrm{d}\log n_b(\omega\!>\!0)/\mathrm{d}\log T$. At low $T$, $\xi(T)$ vanishes in Phase Ia, diverges in Phase II and is a constant in Phase Ib, reflecting $n_b(\omega\!>\!0)\sim T^{\xi}$.
		(c--d) Temperature evolution of the spectra of the self-energies of conduction electrons in (c) Phases I and (d) Phase II. In Phase I a resonance crosses zero energy, while in Phase II the self-energy is always gapped.
		}
\end{figure}

The different behaviors of holon spectral gap in Phase I and II lead to different temperature dependence of uniform channel susceptibility, $\chi_\text{ch}(q\!=\!0)\sim\sum_i \langle \vec{\mathcal{B}}_i \cdot \vec{\mathcal{B}}_j \rangle/\mathcal{V}$, as well as free holon population, $n_b(\omega\!>\!0)\equiv\intoinf{\frac{\mathrm{d}\omega}{2\pi}}{n(\omega)[A_B(\omega)]_{\mathrm{AA}}}$, where $n(\omega)$ is the Bose-Einstein distribution. More importantly, they reveal that Phase~I is further divided into Phase~Ia and Ib, as follows. In Phase Ia where $\gamma\lesssim1$, the free holon population becomes conserved, as seen in the inset of Fig.\,\ref{fig3}(d), while $\chi_\text{ch}$ diverges as $T\to0$ seen in Fig.\,\ref{fig4}(a). In Phase Ib where $1\lesssim\gamma<2$, the free holon population vanishes by a power law $T^\xi$, and $\chi_\text{ch}$ is regular as $T\to0$. In Phase II, the free holon population depletes by $e^{-m_b/T}$ due to the holon gap [inset of Fig.\,\ref{fig3}(d)] , and $\chi_\text{ch}$ stays finite. The power-law exponent of free holon population as a function of $T$ displayed in Figure \ref{fig4}(b) most clearly shows the difference between the three phases. In Phase Ia, the exponent is zero as $n_b(\omega\!>\!0)$ is constant at low $T$. In Phase Ib, the exponent stays constant. In Phase II, the exponent diverges as $T\to0$ due to the exponential decay in $1/T$. (See also Appendix \ref{ss:app:bgap}.)

Another distinct feature between Phase I and II lies in the self-energy of conduction electrons, $N\Sigma_c(k,\omega+i\eta)$. In Phase I, $c$-electron self-energy exhibits two bands of poles that extend throughout the Brillouin zone. In addition, around $\mathrm{K}$ and $\mathrm{K}'$ these poles cross zero energy as $T$ goes down, as plotted in Fig.\,\ref{fig4}(c). On the other hand, $N\Sigma_c$ in Phase II only have poles at $\mathrm{K}$ and $\mathrm{K}'$, and they remained gapped throughout all temperatures, as plotted in Fig.\,\ref{fig4}(d). (See also Appendix \ref{ss:app:sigc} for a Brillouin-zone cut of $N\Sigma_c$.)
Note that no such pole exists in the gapless TR-preserving phase, although a heavily damped pole cannot be ruled out \cite{Hu2021}.

Finally, we note that tuning $t_f$, hence $J_H$, can drive similar transitions in this system with gapped conduction channels, which is discussed in Appendix \ref{ss:app:tf}. If the Haldane mass $m^H_c$ is replaced with a Semenoff mass, $m^S_c$, spinons will also inherit an IR-broken but TR-preserving gap under Kondo interaction. A similar $\gamma$ dependence of holon gap behavior arises as well.

\section{The phase diagram and the interpretation\label{ss:interp}}
Here, we summarize the salient points of the numerical results obtained by the dynamical large-$N$ method. On the honeycomb lattice, we found the following:

(1) \emph{In the presence of TR symmetry}, the ground state is a spin- and channel-symmetry-preserving gapless state, characterized by a line of conformally invariant fixed points parametrized by $\gamma=K/N$. Correlation functions along this line comprise Eqs.\,\eqref{eq:gc} and \eqref{eq:Gb}.

(2) \emph{In the TR-broken regime}, we found three phases, denoted by Ia, Ib and II. In Phase~II bosonic holons are gapped at zero temperature, while in Phase~I the holon gap vanishes as $T$ goes to zero and the holons are gapless. In all phases, fermionic spinons are gapped. However, the spinon spectral dispersion is quite different in the two phases. Phase~I is further divided into Phase~Ia where the holon number becomes conserved at low $T$ and channel susceptibility diverges, and Phase~Ib where the holon number diminishes with decreasing temperature and channel susceptibility saturates.

\subsection{Renormalization group analysis}
The qualitative features of the phase diagram above are present even with a small TR-breaking field. Therefore, a renormalization group (RG) discussion of the nature of these massive phases is in order. For $\gamma>2$ in Phase~II, $\Delta_f<3/2$ as seen in Fig.\,\ref{fig2}(d). This means that a Haldane mass term of the form ${\cal L}={\cal L}_*+m^H_f(\bar \chi_1 \chi_1- \bar \chi_2\chi_2)$ is a relevant perturbation in the RG sense.
Note that a Semenoff mass $m^S_f(\bar \chi_1 \chi_1+ \bar \chi_2\chi_2)$ \cite{Semenoff2012} is forbidden due to the assumed inversion symmetry. 
Indeed, in Phase II, $\braket{\bar \chi_1 \chi_1-\bar \chi_2 \chi_2}\neq 0$. The mass has different signs at the $\mathrm{K}$ and $\mathrm{K}'$, which can be attributed to the Kondo flux repulsion [Fig.\,\ref{fig1}(a)]. In fact in Phase II, $\Delta_f\approx1$ [Fig.\,\ref{fig2}(d)], and the bulk fermion constitutes a noninteracting 2D topological insulator. This characterizes a chiral spin-liquid in which TR-breaking is induced on spinons via the Kondo interaction. 

For $\gamma<2$ such a mass term is irrelevant and the origin of the spinon gap is more subtle. The only relevant interaction in this case is ${\cal L}={\cal L}_*+[V_a(\psi_{1,a\alpha}\dg \chi_{1,\alpha}\dn+\psi_{2,a\alpha}\dg  \chi_{2,\alpha}\dn)+\mathrm{h.c.}]$, with a c-number hybridization $V_a$ which is relevant for all $\gamma$. Indeed, the spinon dispersion in Phase~I can be fitted with a hybridization model between bare $c$- and $f$-electrons (see Appendix \ref{ss:app:hyb}), whereas the TR-broken Haldane model is sufficient to describe the renormalized spinon bands in Phase~II. 

However, such a mass term is forbidden not only due to $\mathrm{SU}(K)$ channel symmetry, but also the internal $\mathrm{U}(1)$ symmetry $\chi \to e^{i\varphi} \chi$ of the Lagrangian.
Similar scenarios have been discussed in the context of symmetric mass generation \cite{You2018,Zeng2022,Wang2022} where a finite $\|V\|$ is maintained but both the phase and direction of the $V$ spinor are randomized.

\subsection{Dynamic mass generation}

An alternative and somewhat more transparent picture is that spin and channel fluctuations can play the role of a glue to bind electrons and spinons. Such fluctuations can be treated by promoting the global charge, spin and channel symmetries to local symmetries and studying the problem of Dirac electrons coupled to nonabelian gauge fields. Gauging these global symmetries also replaces the derivatives on the fields [in Eq.\,\eqref{eq:L*}] by covariant derivatives: $D_c=\partial+A_\text{ex}+A_\text{sp}+A_\text{ch}$ for the electrons, $D_f=\partial+A_\text{in}+A_\text{sp}$ for the spinons, and $D_b=\partial+A_\text{in}-A_\text{ch}$ for the holons. Microscopically, $A_{\rm ch}$ and $A_{\rm sp}$ can be thought of as emerging from spatial/temporal fluctuations of the hybridization (the channel spinor) and magnetization, respectively, which are obtained by decoupling the Kondo interaction in channel/magnetic sectors (see \cite{Wugalter20}).  Henceforth, these gauge symmetries will be denoted by $\mathrm{U}_\text{in}(1)$, $\mathrm{SU}_\text{sp}(N)$, and $\mathrm{SU}_\text{ch}(K)$. 

This enables us to invoke the existing large-$N$ results on 2+1D quantum chromodynamics (QCD3) \cite{Semenoff1994,Appelquist1996,Wijewardhana1997}. Such theories exhibit two phases, a deconfined critical phase equivalent to the Banks-Zaks \cite{Banks1982} conformal fixed point, and a confined phase with chiral symmetry breaking and dynamic mass generation. The latter happens below a critical number of flavors and typically leads to massive particles with conserved particle number splitting off from the continuum. In the following we invoke these results to understand the TR symmetry-broken phase diagram, leaving further analysis of the conformal fixed point to the future \cite{Yang2}.

In 2+1D, abelian gauge fields coupled with many flavors of fermions get screened and are in the deconfining phase \cite{Semenoff1994}. For nonabelian gauge fields with $N_c$ colors coupled to $N_f$ flavor of fermions, in the limit of $N_f\gg N_c$, self-interactions are negligible and they behave like abelian fields, i.e., are in the deconfining phase. On the other hand, in the limit of $N_c\gg N_f$ self-interactions are important and the gauge fields essentially behave as free Yang-Mills which are confining \cite{Semenoff1994}. Since they are coupled to fermions, for $N_c\gg N_f$ there is a tendency to dynamically generate a mass. The critical number of flavors is predicted to be at $\gamma_c=N_f/N_c\to 64/3\pi^2\approx 2.16$.

Since we can vary $N_f/N_c$, both limits are accessible for us. From this discussion, we see that the U$_\text{in}$(1) gauge field coupled to $N$ fermions is always deconfining. But it can also be Higgsed if the bosons, also carrying U$_\text{in}$(1) charge, condense (see below). 
The SU$_\text{sp}$($N$) gauge field has $N$ colors and $K$ flavors. It is in the deconfining phase for $\gamma>\gamma_c$. On the other hand, for $\gamma<\gamma_c$ it is confining and therefore electrons and spinons, which carry corresponding charges, are confined into a bosonic bound state. The role of color and flavor inverts for the SU$_\text{ch}$($K$) gauge field. This one is expected to be confining for $\gamma>\gamma_c'\sim \gamma_c^{-1}$ so that the holons are glued to the electrons. This essentially means the free spinons are Kondo screened channel-symmetrically and the Kondo interaction cannot be decoupled in this limit. For $\gamma<\gamma'_c$ the holons and electron-spinon bound states are deconfined and free to condense. For $\gamma'_c<\gamma<\gamma_c$ electron-spinon bound states have formed, but since they carry channel quantum number, they are confined due to SU$_\text{ch}$($K$) fluctuations. From our numerics, we find $\gamma'_c \sim 1$ rather than $\gamma_c^{-1}\sim 0.463$ predicted by the fully symmetric model.

The spectrum of bosons in Phase I of Fig.\,\ref{fig3}(b) contains a sharp and fully coherent resonance [pole of $G_B(k,\omega+i\eta)$] at low energies which can be attributed to a bound state $\phi\dn_a\sim\psi\dg_{a\alpha}\chi\dn_\alpha$ between conduction electrons and spinons. The bound state is described by ${\cal L}_\phi\sim \bar\phi[D_b^2+m_b^2(T)]\phi$, with its mass approaching zero at zero temperature, $m_b(T\to0)=0$.
This also manifests as the pole in the conduction-electron self-energy $N\Sigma_c$ discussed earlier for $\gamma<2$.

Can the bosonic holon gap closing lead to channel symmetry breaking? Since these bosons are the byproduct of the Hubbard-Stratonovitch decoupling of the Kondo interaction, naively they should not condense as there is no conservation of boson numbers. 
However as discussed in the Results, the number of free holons does become conserved in Phase Ia.
Such an \emph{emergent conservation} of boson number is certainly absent at the UV. At $T=0$, these number conserved bound states reach zero energy and condense in one of the channels, equivalent to spontaneous breaking of the channel symmetry. This is accompanied by the divergence of uniform channel susceptibility in Phase Ia. As $\gamma$ increases beyond $\gamma_c'$, the free holon number is no long conserved, and hence the channel symmetry is restored.

\subsection{Topological order}

\begin{figure}[tp!]
	\includegraphics[width=1\linewidth]{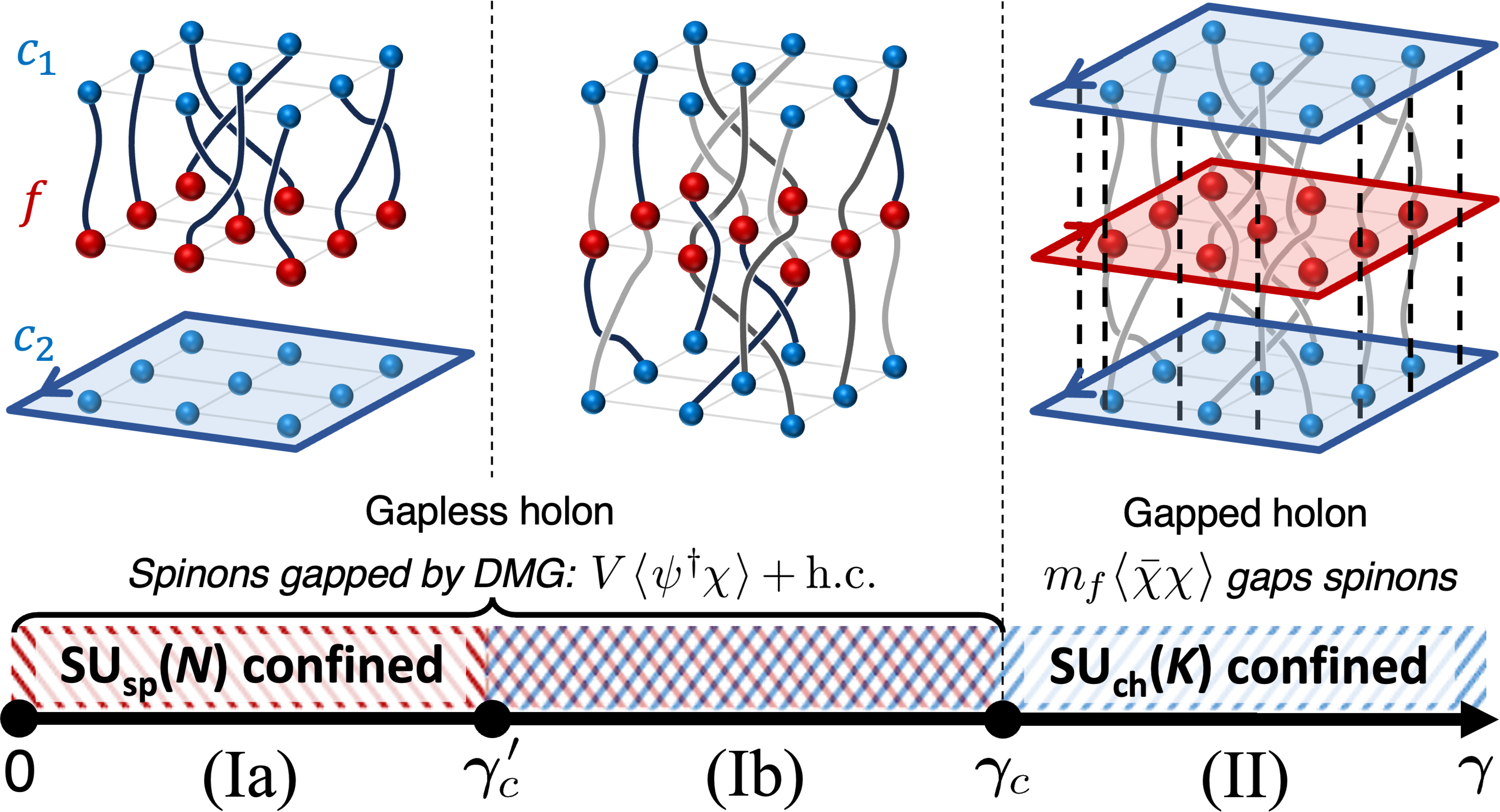}
	\caption{Phase diagram of overscreened Kondo lattices with a Haldane mass in conduction channels. The channel number varies with $\gamma=K/N$ and drives deconfinement transitions of the $\mathrm{SU}_\text{sp}(N)$ and $\mathrm{SU}_\text{ch}(K)$ gauge fields. Phase I has dynamic mass generation (DMG) and by itself is divided into (Ia) a channel ferromagnet where the holon population is conserved at low temperature, and (Ib) a quantum paramagnet where the holon population vanishes by a power law in temperature. Phase II is a fractional Chern insulator with a gapped bulk and counterpropagating edge modes of conduction electrons and spinons, also coupled by Kondo interaction. Numerics suggests $\gamma_c\sim 2$ and $\gamma'_c\sim 1$.}\label{fig5}
\end{figure}

The transition between Phase Ia and Ib is reminiscent of the phase diagram of the NL$\sigma$M describing a channel ferromagnet and a quantum paramagnet with order destroyed by topological defects. The paramagnet is a channel spin liquid, since local moments are screened, and no local order parameter exists. This agreement with Fig.\,\ref{fig1}(b) is not surprising, as in the presence of the gap, fermions can be safely integrated out, and hence the effective interaction of Eq.\,\eqref{eq2} is expected to be valid. These phases are depicted in the phase diagram of Fig.\,\ref{fig5}.

Since in the presence of inversion and time-reversal symmetry the Dirac points in a honeycomb lattice are protected, the only mechanism for the dynamic mass generation is an emergent hybridization. In the entire Phase I, the hybridization is active and the local moments are screened, as evinced by the resonance in the self-energy of conduction electrons  $\Sigma_c$ [see Figs.\,\ref{fig4}(c) and \ref{fig4}(d)]. However, while the spinons respond coherently to the external field due to the Higgs locking between $A_\text{in}$ and $A_\text{ex}$ in the ordered Phase Ia, this coherence is lost in the quantum disordered Phase Ib.

In heavy fermion systems with a FS, the Oshikawa's theorem \cite{Oshikawa00,Senthil2003,Senthil2004} indicates that a Fermi liquid with a small FS can only coexist with topological order. Despite some work in this direction \cite{Kimchi2013,Parameswaran2019} there is no equivalent theorem for our semimetallic system. Nonetheless, heuristically at a Kondo-dominated fixed point the small (large) FS can be associated with inactive (active) hybridization regimes. Therefore, as an analogy to the cases with a FS, the gapped Phase II with inactive hybridization is the only phase with potential for topological order \cite{Wen1990}.

The topological gap in Phase II leads to edge modes composed of counterpropagating electrons $c$ and noninteracting spinons $f$, on an open manifold. These two gapless modes are still coupled via the Kondo interaction $J_K\vec{\cal J}^R\cdot\vec{\cal S}^L$ where ${\cal J}^R_{\alpha\beta}=c\dg_{a\alpha}c\dn_{a\beta}$ and ${\cal S}^L_{\alpha\beta}=f\dg_\alpha f\dn_\beta$ constitute SU$_{K}$($N$) and SU$_{1}$($N$) Kac-Moody currents, respectively. The interaction is marginally relevant and lowers the central charge of the $f$ edge modes from ${\rm c}_\text{UV}/N=1$, due to the c-theorem \cite{ctheorem}. 

In fact, this chiral model is one-half of the 1+1D 2CKL model studied by Andrei and Orignac \cite{Andrei2000,Azaria2000,Azaria1998}. Generalizing their $\mathrm{SU}_\text{sp}(2)\times\mathrm{SU}_\text{ch}(K)$ model to the present $\mathrm{SU}_\text{sp}(N)\times \mathrm{SU}_\text{ch}(K)$ symmetry group, it is natural to expect that the IR edge mode is governed by the \emph{chirally-stabilized} \cite{ADJ1998} fixed point
\begin{equation}
{\cal L}_{\text{Phase II}}\to {\rm SU}_{K-1}{\rm (}N{\rm )}\times\frac{{\rm SU}_{K-1}{\rm (}N{\rm )}\otimes {\rm SU}_{1}{(\rm }N{\rm )}}{{\rm SU}_{K}{\rm (}N{\rm )}},
\end{equation}
in addition to the decoupled charge and channel sectors \cite{Ge22}. This means that Phase II realizes a fractional Chern insulator (FCI) \cite{Regnault2011,Liu2023}, whose edge mode is comprised of a charge mode, an SU(K)$_N$ channel mode, an SU$_{K-1}$($N$) magnetic mode, as well as another anyonic mode from the coset sector. In the large-$N$ limit, this theory leads to a central charge of $\lim_{N\to\infty}{\rm c}_N(\gamma)/N= \gamma/(1+\gamma)$ in addition to fully decoupled conduction electrons \cite{Ge22}.  From bulk-boundary correspondence we expect the bulk to have similar fractionalized description in terms of the gauged Chern-Simons theory.  A detailed computation of electric and thermal Hall conductivities and investigation of other properties of FCI \cite{Sachdev2018}, as well as connections to the order fractionalization \cite{ofc,Tsvelik2022} are beyond scope of the present work and will be reported elsewhere.

\section{Discussion\label{ss:discussion}}

It is worth emphasizing that the only necessary ingredients for the physics observed here are (i) broken TR symmetry to form edge states, and (ii) their further fractionalization by the Andrei-Orignac mechanism. Such topological order is special to MCKL (and therefore different from the one discussed in Ref.\,\onlinecite{Hsieh2017}), but it is not tied to either p-h symmetry or the honeycomb lattice studied here. Kondo flux repulsion generically creates the opposite topology for the spinons to that of electrons, which leads to counterpropagating edge modes of electrons and spinons that are further fractionalized by the multichannel Kondo effect. Remarkably, the physics does not even depend on whether the spin transforms as symmetric or antisymmetric representations. We have used the Schwinger boson representation of the spin and essentially found the same phase diagram apart from a mapping $\gamma\to\gamma^{-1}$, an observation that points towards universality. Lastly, the transition can also be induced by varying $T_K/J_H$ (Appendix \ref{ss:app:tf}).

A remark about magnetic susceptibilities is in order. Overscreened Kondo impurities are critical and unstable against Zeeman splitting \cite{Affleck92}. In Ref.~\onlinecite{Ge22} we showed that this instability carries over to the 1+1D with Schwinger bosons. However, the channel order parameter is a spin-singlet and our Abrikosov fermion calculation (exact in the large-$N$ limit) did not reveal any magnetic ground state in the phase diagram. Fermionic and bosonic representations of the spin are two possible large-$N$ extensions of the SU(2) spin and each capture part of the physics. Since the spin is treated more classically in the Schwinger boson than Abrikosov fermion representation, it is likely that a magnetic ground state is also present, but more simulations are needed to clarify this point.

While the validity of our large-$N$ results for $N=2$ remains to be explored, it is fitting to speculate about experiments. The 2CKL of common spins fall at the border between Phases Ia and Ib $\gamma=2$ and has a chiral Majorana edge state $c_{\text{IR}}=1/2$ (without any superconductivity) at the boundary in the topologically ordered phase. Recent experiment \cite{Zhao2023} and proposals \cite{Kumar2021,Kumar2022,Guerci2023} on realizing Kondo lattices in twisted bilayer graphene and transition metal dichalcogenides suggest that moir\'e systems might be a promising candidate for realization of 2CKL in two spatial dimensions.

In conclusion, by including spatial fluctuations in a MCKL in two spatial dimensions, we have shown that the channel symmetry is preserved as opposed to the findings of single-site DMFT and static mean-field calculations. The phase diagram Fig.\,\ref{fig1}(b) of NL$\sigma$M does not apply to a Kondo lattice below the upper critical dimension, as the gapless fermions drive the system toward a conformally invariant fixed point. We have characterized this fixed point in the large-$N$ limit by finding the critical exponents using a scaling ansatz. Breaking TR symmetry with the addition of a Haldane mass to conduction electrons gaps out spinons and holons by the resonant RKKY amplification. In this case, however, we recover the phase diagram of Fig.\,\ref{fig1}(b) so that in one phase both spinons and holons are gapped, while in the other phase an emergent stabilization of holon population and a temperature-dependent gap indicate a gapless ground state for the holons and the spontaneous breaking of channel symmetry similar to the NL$\sigma$M description. We discussed the role of gauge fields and argued that the gapped phase realizes a FCI, with edge modes governed by the theory of Andrei and Orignac \cite{Andrei2000}. The fact that the same edge theory appears in certain nonabelian fractional quantum Hall states, points to a deep and fascinating connection between the latter and MCKL, which remains to be explored.

\begin{acknowledgments}
	It is a pleasure to acknowledge fruitful discussions with N.\ Andrei, P.\ Coleman, S.~Raghu and R.\ Wijewardhana. Part of this work was performed at Aspen Center for Physics, which is supported by NSF Grant No.\ PHY-1607611. Computations for this research were performed on the Advanced Research Computing Cluster at the University of Cincinnati, and the Ohio Supercomputer Center, Ohio, USA.
\end{acknowledgments}

\appendix

\section{Summary of 1+1D overscreened Kondo lattice results \label{ss:app:1d}}

The two-channel Kondo lattice was studied in 1+1D using the Schwinger boson representation of the spins in Ref.\,\onlinecite{Ge22}. In this work, we use the Abrikosov fermion representation instead. In the large-$N$ limit they are closely related by a duality, discussed at the end of this section. Here we summarize the essential results in 1+1D.

We consider the same Kondo lattice model in Eq.\,\eqref{eqH} on a simple 1+1D lattice. The spins are represented by the Abrikosov fermions, also referred to as $f$-electrons or spinons. The model becomes essentially identical to that in Sec.\,\ref{ss:method}, without the sublattice structure:
\begin{eqnarray}
\label{eq:H1D}
	H & = &  H_c+H_f+H_K,  \nonumber\\
	H_c & = &  \sum_k (- 2 t_c \cos k - \mu_c) c^{\dag}_{k a \alpha} c\dn_{k a \alpha},  \nonumber\\
	H_f & = & \sum_i \frac{N | t_f |^2}{J_H} + \sum_k (- 2 t_f \cos k - \mu_f) f_{k \alpha}^{\dag} f\dn_{k \alpha}, \nonumber\\
	H_K & = &  \sum_i \frac{| \bar{b}_{i a} b_{i a} |}{J_K} + \frac{1}{\sqrt{N}} \sum_i  \left( b\dn_{i a} f_{i \alpha}^{\dag} c\dn_{i a \alpha} + \text{h.c.} \right)\!. \quad 
\end{eqnarray}
The spin index is $\alpha=1,\dots,N$, and the channel index is $a=1,\dots,K$, with $\gamma=K/N\sim\mathrm{O}(1)$. Note that at UV, the holons $b, \bar{b}$ are Hubbard-Stratonivich fields with no dynamics. Hence, holons have no time derivative in the action [Eq.\,\eqref{eq3}]. In addition, the NN hopping amplitude is conventionally positive for $t_c$, while it is negative for $t_f$ due to the decoupling of the antiferromagnetic Heisenberg interaction. Setting $\mu_f=0$ gives the self-conjugate representation for the spins, as is used for 2+1D, while p-h symmetry holds if $\mu_c=0$ as well. At large $N$, the system can be exactly solve to $\mathrm{O}(1)$ with the self-consistent equations \eqref{eqself}--\eqref{eqdyson}. 

At low temperature, the 1+1D overscreened Kondo lattice model \eqref{eq:H1D} exhibits the emergent dispersion of $f$-electrons or spinons, and conformal ground states, while the magnetic and channel susceptibilies of the system are defined by the scaling exponents. Below we discuss these features in detail.

\subsection{Emergent dispersion of spinons}

Even when $t_f = 0$ at UV, an emergent dispersion for spinons will be generated at low $T$ via a resonant RKKY amplification mediated by Kondo interaction {\cite{Ge22}}. This emergent dispersion will break the Galilean boost symmetry $(f_j, b_j) \rightarrow e^{i k j} (f_j, b_j)$. In the numerics, this manifests as an instability toward dispersion at $t_f = 0$ with an arbitrary small spatial correlation added to the spinon self-energy. This also entails that an infinitesimal $J_H$ can lead to an amplified dispersion at low $T$ in the overscreened Kondo lattice. Consequently, when $t_f$ is small, the effective bandwidth of spinon self-energy at zero frequency can overwhelm the bare dispersion, i.e.\ $\max [\Sigma_f (k,\omega = 0)] - \min[\Sigma_f (k,\omega = 0)] \gg 2|t_f|$. The same effect is observed in 2+1D, shown in Fig.\,\ref{fig:app:Eeff}(a) and \ref{fig:app:Eeff}(b).

\subsection{The conformal fixed point in 1+1D \label{ss:app:1dCFT}}

\begin{figure}
	\includegraphics[width=0.55\linewidth]{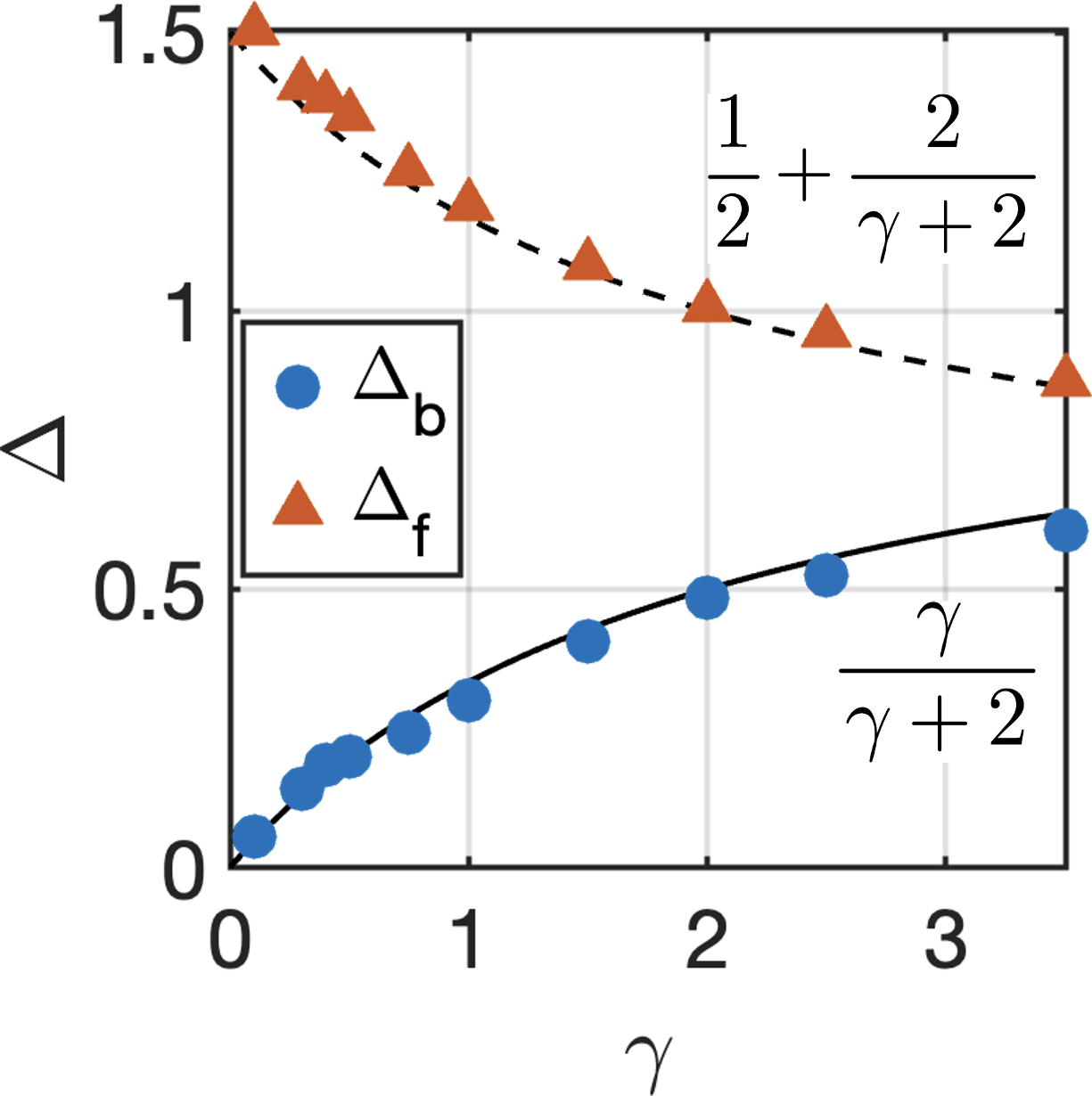}
	\caption{\label{fig:app:1Dexp} Scaling exponents for the 1+1D overscreened Kondo lattice using large-$N$ Abrikosov fermion representation. Solid and dashed curves are the conformal solution exponents. }
\end{figure}

Due to overscreening, $f$ and $b$ often remains critical at low $T$ in the absence of spontaneous symmetry breaking. When the Fermi momenta $k_F$ of $c$ and $f$ are nesting, i.e.\ $n_c+n_f=1$, the system flows to a conformal fixed point at low $T$. At this fixed point, $f$-electrons host chiral linear modes at $\pm k_F$, while the holon host a pair of linear dispersing modes at $k=0$ with the same group velocity $v$. Let $z=v\tau+ix$, the conformal solution in 1+1D consists of the low-energy modes at these critical momenta:
\begin{gather}
G_{f,R} = \alpha_f z|z|^{-2\Delta_f-1}, \quad G_{f,L} = \alpha_f \bar{z}|z|^{-2\Delta_f-1}, \nonumber \\
G_b =\alpha_b |z|^{-2\Delta_b}, 
\end{gather}
where $G_{f,L/R}$ are the left- and right-propagating $f$-modes at $\pm k_F$, respectively. The scaling exponents $\Delta_{f,b}$ can be extracted numerically from the scaling collapse of spectral functions at low $T$. They agree with the analytical solution obtained from substituting the conformal ansatz into the self-energy equations and set $G(k,\omega)\Sigma(k,\omega)=-1$.
The 1+1D solutions are 
\begin{equation}
\label{eq:1dexp}
	\Delta_f=\frac12+\frac{2}{\gamma+2}, \qquad \Delta_b=\frac{\gamma}{\gamma+2},
\end{equation}
as shown in Fig.\,\ref{fig:app:1Dexp}.

In the limit $\gamma\to0$, the Kondo lattice becomes perfectly screened. This is reflected in a constant hybridization owing to $\Delta_b=0$. Meanwhile, the local spinon spectral function vanishes by $A_f(\omega)\sim\omega^2$ near zero frequency. The perfect-screening limit
\begin{equation}
	\label{eq:Db@0}
	\lim_{\gamma\to0}\Delta_b=0,
\end{equation}
holds for both the overscreened Kondo impurity and 1+1D Kondo lattice. We expect it to be true in all dimensions.

\subsection{Susceptibilities \label{ss:app:sus}}
The magnetic and channel susceptibilities in the dynamical large-$N$ model are given by, with $\vec{r}\equiv(\mathbf{r},\tau)$,
\begin{eqnarray}
	\chi_{\text{m}}(\vec{r}\,) &=& G_f(\vec{r}\,) G_f(-\vec{r}\,), \\
	\chi_{\text{ch}}(\vec{r}\,) &=& J^{-4}_K [G_b(\vec{r}\,)\!+\!J_K \delta(\vec{r}\,)] [G_b(-\vec{r}\,)\!+\!J_K \delta(\vec{r}\,) ]. \quad
\end{eqnarray}
In particular, $\chi_{\text{ch}}$ measures the susceptibility to asymmetries in the Kondo coupling to different channels. At the conformal fixed point, the scaling part of $\chi_{\text{m}}$ and $\chi_{\text{ch}}$ obey power laws set by the conformal exponents, i.e.\ Eq.\,\eqref{eq:1dexp}.

At low $T$, the static $\chi_{\text{m}}$'s, both the local and the uniform susceptibilities, are nondivergent in 1+1D for all $\gamma$. On the other hand, $\chi_{\text{ch}}(\mathbf{r}=0,\omega=0)$ is divergent for $4\Delta_b-1<0$, whereas $\chi_{\text{ch}}(\mathbf{k}=0,\omega=0)$ is divergent for $4\Delta_b-2<0$ in 1+1D. This entails that at low $T$, the overscreened Kondo lattice model with the Abrikosov fermion representation is divergently susceptible to channel asymmetry perturbations for small $\gamma$, due to Eq.\,\eqref{eq:Db@0} in all dimensions. Despite the divergence, the conformal fixed point is channel symmetric, and there is no spontaneous channel symmetry breaking.

\subsection{Duality between Abrikosov fermion and Schwinger boson representations}

In the dynamical large-$N$ method, Abrikosov fermion and Schwinger boson representation of the spins lead to the low energy effective theories for the overscreened Kondo impurity or lattice that are dual to each other at the conformal fixed points. In the latter case, the spinons are Schwinger bosons, also denoted by $b$, whereas the holons are fermionic fields $\chi$~\cite{Komijani18,Ge22}. The duality is
\begin{equation}
	f \leftrightarrow \chi, \quad b \leftrightarrow b , \quad \gamma \leftrightarrow 1/\gamma,
\end{equation}
and the $c$-electron remains unchanged.

\section{Numerical method for the dynamical large-$N$ Kondo lattice model \label{ss:app:numerics}}
\subsection{Numerical self-consistent equations}
The Dyson equations for spinons and holons together with the self-energy equations in Eqs.\,\eqref{eqself} constitute a set of self-consistent equations. In the present case of the Kondo lattice with the large-$N$ Abrikosov fermion representation of the spins, they are numerically solved to $\mathrm{O}(1)$ by essentially the same method applied to the Schwinger boson representations in Ref.\,\onlinecite{Ge22}. We derive the details of self-consistent equations below. For this discussion we do not use the spinor notation introduced in Eq.\,\eqref{eq:spinor}. A Green function $(G_c)_{mn}$ is associated with a field $c$ propagating from sublattice site $n$ to $m$. They are related to the spinorial versions by $g_c=G_C \sigma^z$, $G_f=G_F \sigma^z$, and $G_b=G_B$. 

The self-energies for a lattice displacement-imaginary time $\vec{r}=(i,j,\tau)$ are
\begin{eqnarray}
	\left[ \Sigma_f (\vec{r}\,) \right]_{mn} & = & - \gamma [g_c(\vec{r}\,)]_{mn}  [G_b (\vec{r}\,)]_{mn},\\
	\left[ \Sigma_b (\vec{r}\,) \right]_{mn} & = & [g_c (-\vec{r}\,)]_{nm}  [G_f (\vec{r}\,)]_{mn}, \\
	N \left[\Sigma_c (\vec{r}\,)\right]_{mn} & = & - [G_b (-\vec{r}\,)]_{nm}  [G_f (\vec{r}\,)]_{mn} , 
\end{eqnarray}
Note the sublattice transpose for backward propagation ($- \vec{r}$\,). These equations are solved using spectral functions of the Green's functions and self-energies in real frequency-momentum space, dropping $\Sigma_c$ at $\mathrm{O}(1)$. Self-energies are obtained from Hilbert transforms of $A_\Sigma(\omega) \coloneqq i \Sigma(\omega \! + \! i\eta) - i\Sigma(\omega \! - \! i\eta)$, where $\eta \! \to \! 0^+$. They are
\begin{widetext}
	\begin{eqnarray}
		\left[ {A_{\Sigma_f}} (\mathbf{k}, \omega) \right]_{mn} & = & - \frac{\gamma}{\mathcal{V}}  \sum_{\mathbf{p}} \int \frac{\mathrm{d} \nu}{2 \pi}  [A_c (\mathbf{p}, \nu)]_{mn} [A_b (\mathbf{k}-\mathbf{p}, \omega - \nu)]_{mn}  [f (\nu) + n_B (\nu - \omega)], \label{eq:app:ASf} \\
		\left[ {A_{\Sigma_b}} (\mathbf{k}, \omega) \right]_{mn} & = & \frac{1}{\mathcal{V}} \sum_{\mathbf{p}} \int \frac{\mathrm{d} \nu}{2 \pi}  [A_c (\mathbf{p}, \nu)]_{nm} [A_f (\mathbf{k}+\mathbf{p}, \omega + \nu)]_{mn}  [f (\nu) - f (\omega + \nu)],  \label{eq:app:ASb} \\
		N [A_{\Sigma_c} (\mathbf{k}, \omega)]_{m n} & = & \frac{1}{\mathcal{V}} \sum_\mathbf{p} \int \frac{\mathrm{d} \nu}{2 \pi}  [A_b (\mathbf{p}, \nu)]_{n m} [A_f (\mathbf{k}+\mathbf{p}, \omega+\nu)]_{m n}  [f (\omega+\nu) + n(\nu)] ,
	\end{eqnarray}
	where $f$ and $n$ are respectively Fermi-Dirac and Bose-Einstein distributions, and $\mathcal{V}$ is the system size. Since $c$-electrons remain bare at $\mathrm{O}(1)$, we can use the explicit form of $A_c$. Denote the Pauli matrices by $\vec{\sigma} \coloneqq (\sigma^x, \sigma^y, \sigma^z)$. For $H_c (\mathbf{k}) = \vec{\epsilon}_{\mathbf{k}} \cdot \vec{\sigma} - \mu_{c\mathbf{k}}  \mathbb{1}$. $g_c^{- 1} (\mathbf{k}, \mathrm{z}) = (\mathrm{z}+ \mu_{c\mathbf{k}})  \mathbb{1} - \vec{\epsilon}_{\mathbf{k}} \cdot \vec{\sigma}$. When $\epsilon_{\mathbf{k}} \coloneqq \| \epsilon_{\mathbf{k}} \| \neq 0$, the spectral function of $c$-electron  is
	\begin{equation}
		A_c (\mathbf{k}, \omega) = \pi \left[ \left( \mathbb{1} + \frac{\vec{\epsilon}_{\mathbf{k}} \cdot \vec{\sigma}}{\epsilon_{\mathbf{k}}} \right) \delta (\omega + \mu_{c\mathbf{k}} - \epsilon_{\mathbf{k}}) + \left( \mathbb{1} - \frac{\vec{\epsilon}_{\mathbf{k}} \cdot \vec{\sigma}}{\epsilon_{\mathbf{k}}} \right) \delta (\omega +\mu_{c\mathbf{k}} + \epsilon_{\mathbf{k}}) \right].
	\end{equation}
	When $\epsilon_{\mathbf{k}} = 0$, it reduces to $A_c (\mathbf{k}, \omega) = 2 \pi \delta (\omega + \mu_{c\mathbf{k}})  \mathbb{1}$. Substituting $A_c$ into Eqs.\eqref{eq:app:ASf} and \eqref{eq:app:ASb} gives
	\begin{eqnarray}
		\left[ {A_{\Sigma_f}} (\mathbf{k}, \omega) \right]_{mn} & = & - \frac{\gamma}{2\mathcal{V}}  \sum_{\mathbf{p}} \sum_{s = \pm} \left[ \mathbb{1} + \frac{\vec{\epsilon}_{\mathbf{p}} \cdot \vec{\sigma}}{s \epsilon_{\mathbf{p}}} \right]_{mn} [A_b (\mathbf{k}-\mathbf{p}, \omega - s \epsilon_{\mathbf{p}})]_{mn}  [f (s \epsilon_{\mathbf{p}}) + n_B (s \epsilon_{\mathbf{p}} - \omega)], \label{eq:app:ASfk} \\
		\left[ {A_{\Sigma_b}} (\mathbf{k}, \omega) \right]_{mn} & = & \frac{1}{2\mathcal{V}}  \sum_{\mathbf{p}} \sum_{s = \pm} \left[ \mathbb{1} + \frac{\vec{\epsilon}_{\mathbf{p}} \cdot \vec{\sigma}}{s \epsilon_{\mathbf{p}}} \right]_{nm} [A_f (\mathbf{k}+\mathbf{p}, \omega + s \epsilon_{\mathbf{p}})]_{mn}  [f (s \epsilon_{\mathbf{p}}) - f (\omega + s \epsilon_{\mathbf{p}})]. \label{eq:app:ASbk}
	\end{eqnarray}
\end{widetext}
The retarded or advanced self-energies are obtained from $A_\Sigma$'s with a Hilbert transform, $\Sigma(\omega\pm i \eta)=-\frac{1}{2}\mathfrak{H}[A_\Sigma](\omega) \mp \frac{i}{2} A_\Sigma(\omega)$. Next we need the Dyson equations
\begin{eqnarray}
	G_f^{-1}( \mathbf{k},{\rm z}) & = & {\rm z}\mathbb{1}+\hat{\mu}_f-H_f( \mathbf{k})-\Sigma_f( \mathbf{k},{\rm z}), \label{eq:app:Gf}\\
	G_b^{-1}( \mathbf{k},{\rm z}) & = &-\hat{J}_{K}^{-1}-\Sigma_b( \mathbf{k},{\rm z}).\label{eq:app:Gb}
\end{eqnarray}
Here $\hat{\mu}_f$ and $\hat{J}_K$ are diagonal, but may not be proportional to $\mathbb{1}$, e.g., when inversion symmetry is broken. In the main text, $\hat{\mu}_f=\mu_f \mathbb{1}$ and $\hat{J}_K=J_K \mathbb{1}$ always.

The system is solved by iterating through Eqs. \eqref{eq:app:ASfk}--\eqref{eq:app:Gb}. At the end of each iteration, one needs to adjust $\mu_f$ to satisfy the constraint $2S/N=\langle f^\dagger f \rangle$ on both sublattices. As outlined in Ref.\,\onlinecite{Ge22}, we start at a high temperature, where we run the self-consistency iterations until $G$'s converge, and then repeat at lower temperatures.

The constraint simplifies when the entire system has particle-hole symmetry on all sublattice sites. This occurs when the spin size is $S/N=1/2$, while $H_c$ and $H_f$ have p-h and inversion symmetries. Then $\mu_f=\mu_c=0$ always. This condition is used unless otherwise specified.

Note that similar to the case with Schwinger bosons~\cite{Ge22}, the diagonal part of the holon Green's function $G_b(\omega \pm i\eta)_{nn}$ is not strictly casual, i.e., $[G_b(\omega \pm i\eta)]_{nn}+1/2\{\mathfrak{H}[A_b](\omega) \pm i A_b(\omega)\}_{nn}=-J_K \neq 0$, and neither is $\Sigma_f$. However, the difference is always a real constant. It will be automatically absorbed in $\mu_f$ when applying the constraint on $G_f$.

\subsection{Reducing the Brillouin zone to the fundamental domain \label{ss:app:numsym}}

\begin{figure}
	\includegraphics[width=1\linewidth]{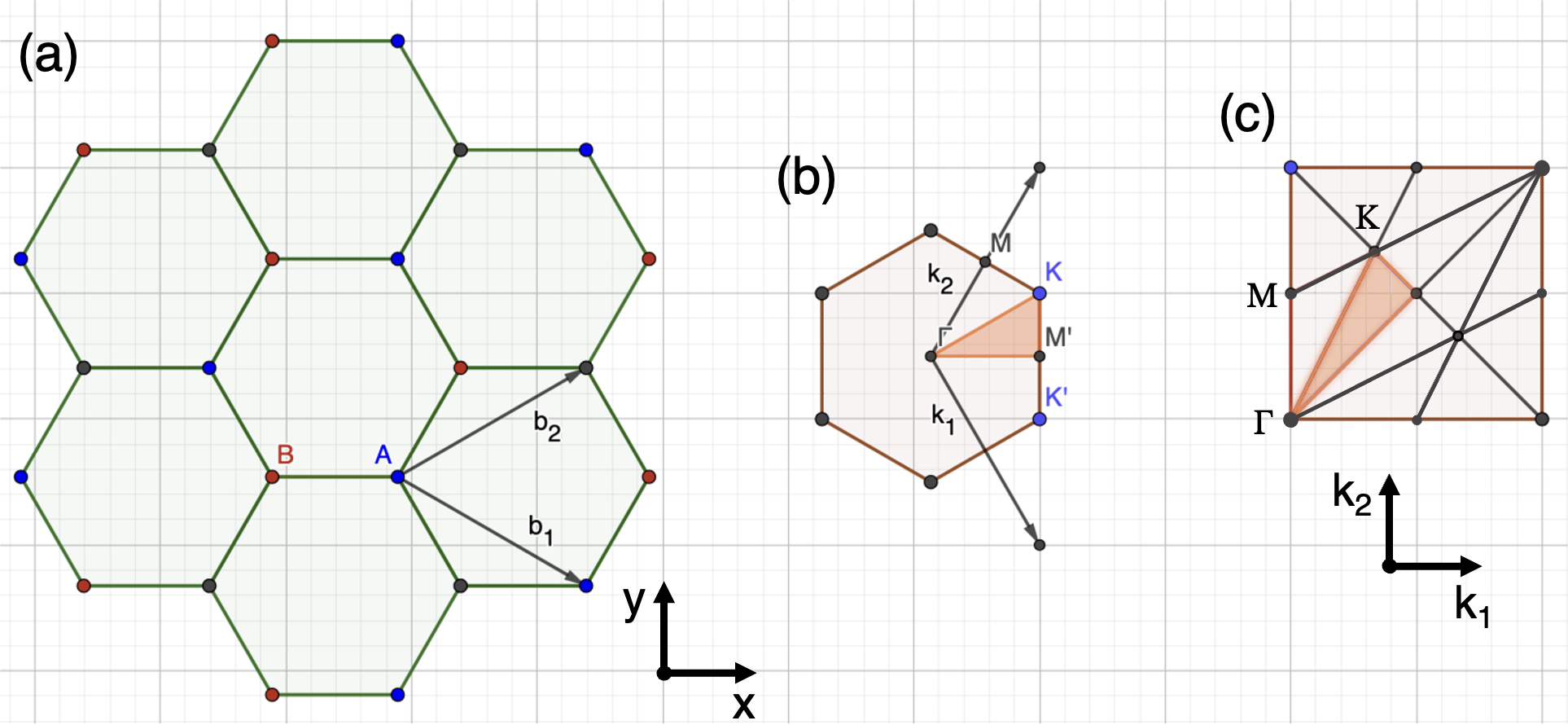}
	\caption{\label{fig:app:1BZ}(a)~Honeycomb lattice with the unit cell A-B sublattice sites labeled, as well as the Bravais vectors $\mathbf{b}_{1,2}$. (b)~First Brillouin zone in the $(k_x,k_y)$-coordinate. The reciprocal Bravais vectors $\mathbf{k}_1$ and $\mathbf{k}_2$ are shown. The fundamental domain is highlighted in orange. (c)~First Brillouin zone in the $(k_1,k_2)$-coordinate, divided into the 12 images of the fundamental domain.}
\end{figure}

\begin{table*}
	\caption{\label{tbl:app:symop}Momentum space maps and sublattice-basis representations of the crystalline symmetries of the honeycomb lattice, and of some internal symmetries: time-reversal $\mathcal{T}$, sublattice $\mathcal{S}$, and particle-hole $\mathcal{C}$. Their maps $R:(k_1,k_2)\to(k_1',k_2')$ are given in the reciprocal Bravais vector coordinate of Fig.\,\ref{fig:app:1BZ}. Their unitary representations $U'_R$ in the sublattice basis are given in the $C_3$-invariant Bloch basis gauge choice. These symmetries do not act on spins.}
	\begin{ruledtabular}
		\begin{tabular}{c c c c c c c c c}
			& $C_3$ & $C_2$ & $C_6$ & $M_x$ & $M_y$ & $\mathcal{T}$ & $\mathcal{S}$ & $\mathcal{C}$ \\ \hline
			$R\mathbf{k}$ & $(- k_2, k_1 - k_2)$  & $(- k_1, - k_2)$ & $(k_1 - k_2, k_1)$ & $(k_2, k_1)$ & $(- k_2, - k_1)$ & $(- k_1, - k_2)$ & $(k_1, k_2)$ & $(- k_1, - k_2)$ \\ \hline
			$U'_R$ & $\mathbb{1}$ & $\sigma^x$ & $\sigma^x$ & $\mathbb{1}$ & $\sigma^x$ & $\mathcal{K}$ & $\sigma^z$ & $\sigma^z \mathcal{K}$ 
		\end{tabular}
	\end{ruledtabular}
\end{table*}

A full calculation throughout the first Brillouin zone (1BZ) is expensive. We compute only $\mathbf{k}$-points in the fundamental domain of our Lagrangian on the 1BZ~\cite{Monkhorst1976,kruthoff_topological_2017}, to which the Green's functions on the rest of 1BZ are related by symmetry transformations. With full crystalline symmetries of the honeycomb lattice $C_{6v}$, the number of $\mathbf{k}$-points needed is reduced by a factor of 12 in large systems. Below we discuss this procedure in detail. In addition to the momentum-space symmetries, we note that the computation load can be further reduced by utilizing the p-h symmetry and any remaining degeneracies for Eqs.\,\eqref{eq:app:ASfk}--\eqref{eq:app:ASbk}.

The honeycomb lattice and its 1BZ is drawn in Fig.\,\ref{fig:app:1BZ}, with unit cell defined by a pair of AB sublattice sites along $x$, separated by $\mathbf{d}_0 \coloneqq \overrightarrow{\mathrm{AB}}$. Before writing down the symmetry operators we need to pick the gauge(s) for the Bloch basis we use. The gauge most convenient to the self-consistent equations is the reciprocal-lattice periodic gauge,
\begin{equation} \left(\left|u_\mathbf{k}\right\rangle_\mathrm{A},\left|u_\mathbf{k}\right\rangle_\mathrm{B}\right)=\frac{1}{\sqrt{\mathcal{V}}} 
	\sum_\mathbf{k} e^{-i \mathbf{k} \cdot \mathbf{r}_j} \left(\left|u_j \right\rangle_\mathrm{A},\left|u_j \right\rangle_\mathrm{B}\right).
\end{equation}
Other gauges may require a twist at the boundary of 1BZ in the sums of Eqs.\,\eqref{eq:app:ASfk}--\eqref{eq:app:ASbk}. Convenient for our symmetry discussion is another gauge choice,
\begin{equation} \left(\left|u'_\mathbf{k}\right\rangle_\mathrm{A},\left|u'_\mathbf{k}\right\rangle_\mathrm{B}\right)=\frac{1}{\sqrt{\mathcal{V}}} 
	\sum_\mathbf{k} e^{-i \mathbf{k} \cdot \mathbf{r}_j} \left(\left|u_j \right\rangle_\mathrm{A},e^{-i \mathbf{k} \cdot \mathbf{d}_0} \left|u_j\right\rangle_\mathrm{B}\right).
\end{equation}
We refer to it as the $C_3$-invariant gauge, and denote quantities in this gauge by the \emph{prime}. Hamiltonians and one-particle Green's functions in these two gauges are related by the gauge transformation $\Lambda(\mathbf{k})$, expressed below in the $(k_1,k_2)$-coordinate shown in Fig.\,\ref{fig:app:1BZ}(c), with $b$ being the length of a real-space Bravais vector.
\begin{eqnarray}
	[ G, H ] (\mathbf{k}) & = & \Lambda (\mathbf{k})  [ G', H' ] (\mathbf{k}) \Lambda^{\dag} (\mathbf{k}) \\
	\Lambda (\mathbf{k}) & = & \begin{pmatrix}
		1 & \\
		& e^{i \mathbf{k} \cdot \mathbf{d}_0 }
	\end{pmatrix} = \begin{pmatrix}
		1 & \\
		& e^{- i (k_1 + k_2) b / 3}
	\end{pmatrix} . \nonumber 
\end{eqnarray}

A symmetry operation $R$ is represented in reciprocal space by unitary transformations and $\mathbf{k}$-mappings, 
\begin{equation}
	U'_R (\mathbf{k}) H' (\mathbf{k}) U'^{\dagger}_R (\mathbf{k}) = H' (R\mathbf{k}).
\end{equation}
The $C_{6v}$ group of the lattice has rotation symmetries $C_2$, $C_3$ and $C_6$, as well as in-plane reflection symmetries $M_x$ and $M_y$. They are listed in Table.\,\ref{tbl:app:symop} along with the internal symmetries. Note that the time-reversal symmetry $\mathcal{T}$ we use here does not act on the spins. To go back to the periodic gauge, we use
\begin{equation}
	U_R (\mathbf{k}) = \Lambda (R\mathbf{k}) U_R' (\mathbf{k}) \Lambda^{\dag} (\mathbf{k}) .
\end{equation}
For example, in the periodic gauge $C_3 \circeq  \begin{psmallmatrix}
	1 & \\
	& e^{i k_2 b}
\end{psmallmatrix}$. 

The action or Hamiltonian may break some of the bare lattice symmetries. Accordingly, we pick different set of generators for different cases.
\begin{enumerate}
	\item With only real NN hopping, Kondo interaction and uniform chemical potentials, we pick $\{ C_3, C_2, M_x \}$ as generators.
	\item In the presence of the Haldane mass, $M_{x}$, $M_y$ and $\mathcal{T}$ are broken. We pick $\{ C_3, C_2, M_{y} \mathcal{T} \}$ as generators.
	\item In the presence of sublattice staggering or the Semenoff mass, $M_y$ and $C_2$ are broken. We pick $\{ C_3, \mathcal{T},M_x \}$ as generators.
\end{enumerate}
The fundamental domain we use is the triangle ${\bigtriangleup}{\Gamma}\mathrm{KM}'$ in Fig.\,\ref{fig:app:1BZ}(b). The symmetry generators listed above can be used to span the full 1BZ. Table.\,\ref{tbl:app:gen1BZ} details the fundamental domain and its generators for Case 1.

\begin{table}
	\caption{\label{tbl:app:gen1BZ}Image generators of the fundamental domain for honeycomb lattice with nearest-neighbor hopping only. $\overline{\Gamma \mathrm{M}'}$ denotes line segments between $\Gamma$ and $\mathrm{M}'$ excluding endpoints, ${\bigtriangleup}{\Gamma}\mathrm{KM}'$ denotes the triangular region defined by the three points excluding edges, etc. }	
	\begin{ruledtabular}
		\begin{tabular}{r|l|l|l|l|l|l|l}
			$k$-region & $\Gamma$ & $\mathrm{K}$ & $\mathrm{M}'$ & $\overline{\Gamma \mathrm{M}'}$ & $\overline{\mathrm{K} \mathrm{M}'}$ & $\overline{\Gamma \mathrm{K}}$ & ${\bigtriangleup}{\Gamma}\mathrm{KM}'$ \\
			Generators & $\emptyset$ & $M_x$ & $C_3$ & $C_3, C_2$ & $C_3, M_x$ & $C_3, M_x$ & $C_3, M_x, C_2$
		\end{tabular}
	\end{ruledtabular}
\end{table}

Note that due to hermicity, $A_{\mathrm{AB}}=A_{\mathrm{BA}}^{\ast}$ always. With only real hopping, the system has the $C_2 \mathcal{T}$-symmetry, which does not change $\mathbf{k}$, and is represented by $C_2\mathcal{T} \circeq \sigma^x \mathcal{K}$ where $\mathcal{K}$ is the complex conjugation operator. It ensures $A_{\mathrm{AA}} {= A_{\mathrm{BB}}} $. Together with $C_2$, it gives $[A (- k)]_{\mathrm{AA}} = [A (k)]_{\mathrm{BB}}$.

Finally, with NN hopping only, there exists a subextensive degeneracy usually present in the $(k_1, k_2)$-coordinate, on $\overline{\mathrm{M}\mathrm{M}'}$ lines. That is, $\epsilon_\mathbf{k} = | 1 + e^{i k_1 b} + e^{i k_2 b} |$ is constant when $k_1 b= \pi$ or $k_2 b = \pi$.

\section{Overscreened Kondo conformal fixed point with gapless conduction channels \label{ss:app:cft}}

In this section, we present further calculations and numerical results to supplement the discussion about the overscreened Kondo conformal fixed point on a Kondo honeycomb lattice with gapless conduction electrons in Sec.\,\ref{ss:gapless}.

\subsection{Conformal ansatz on the honeycomb lattice \label{ss:app:conformal}}
Here we show detailed calculations using the conformal ansatz in Eqs.\,(\ref{eq:Gf}). They solve the self-consistent equations with the condition that at the Kondo fixed point, so that $G_B\Sigma_B=-\mathbb{1}$ in frequency-momentum space, and similarily for the $f$-electron at $\mathrm{K}$ and $\mathrm{K}'$, $G_{F,\mathrm{K}}\Sigma_{F,\mathrm{K}}=G_{F,\mathrm{K}'}\Sigma_{F,\mathrm{K}'}=-\mathbb{1}$. At the low temperature fixed point, we have equal group velocities $v_f=v_b=v$. We further assume that $v_c=v$ to simplify our analysis, similar to the study in 1+1D (Appendix\,\ref{ss:app:1d}).

Denote $\vec{r} = (x, y, v \tau)$, $\cancel{r} \coloneqq r^i \sigma^i = \begin{psmallmatrix}
	v \tau & x - i y\\
	x + i y & - v \tau
\end{psmallmatrix}$, and $\bcancel{r} \coloneqq \cancel{r}^{\ast} = \cancel{r}^{\,\scriptscriptstyle{\mathrm{T}}}$. 
Since $\delta(\vec{r}\,)=-\partial^2 \frac{1}{r}=-\cancel{\partial}\cancel{\partial} \frac{1}{r}=\cancel{\partial} \frac{\slashed{r}}{r^3}$,
the Green's function for $c$-electrons in the long wavelength limit is
\begin{equation}
	g_C (\vec{r}\,) = \frac{1}{4 \pi r^3}  \Big( e^{- i \mathbf{K}\mathbf{r}}  \cancel{r} + e^{ -i \mathbf{K}'\mathbf{r}} \bcancel{r} \Big).
\end{equation}
We denote by bold symbols the 2D $(x, y)$- or $(k_x, k_y)$-vectors. Also note that $\mathbf{K} = - \mathbf{K}'$, up to reciprocal-lattice translations.

On the honeycomb lattice with real NN hopping only, we use the ansatzes in Eqs.\,(\ref{eq:Gf}), up to $\sigma^z$,
\begin{eqnarray}
	G_F (\vec{r}\,) & = &  \frac{\alpha_f}{4 \pi r^{2 \Delta_f + 1}}  \Big( e^{- i \mathbf{K}\mathbf{r}} U \cancel{r} U^{\dagger} + e^{i \mathbf{K} \mathbf{r}} U \bcancel{r} U^{\dagger} \Big), \nonumber\\
	G_B (\vec{r}\,) & = &   \frac{\alpha_b}{4\pi r^{2 \Delta_b}}\mathcal{P}_b,  \label{eq:appGf}
\end{eqnarray}
Here, $U \coloneqq \exp (i \sigma^z \vartheta / 2)$, and $\mathcal{P}_b$ is a projection matrix in the sublattice basis:
\begin{equation}
	\mathcal{P}_b = U\frac{1}{2} \mat{
		1 & 1\\
		1 & 1}
	U^{\dagger} =\frac{1}{2} \begin{pmatrix}
		1 & e^{- i \vartheta}\\
		e^{ i \vartheta} & 1
	\end{pmatrix}.
\end{equation}
The angle $\vartheta$ captures the phase offset between Dirac cones of $c$- and $f$-electrons. With no loss of generality, we set $\vartheta=0$, hence $U=1$.

Let $\vec{q}\coloneqq(k_x,k_y,\omega_m/v)$. The Fourier transform for $G_B$ is
\begin{eqnarray}
	G_B (\vec{q}\,) & \equiv & \frac{\alpha_b}{4\pi} \mathcal{P}_b \int \mathrm{d}^3r \,r^{- 2 \Delta_b}
	\exp (i \vec{q} \cdot \vec{r}\,) \nonumber\\
	& = & \frac{\alpha_b}{4\pi} \mathcal{P}_b q^{2 \Delta_b - 3} \! \int_0^{\infty} \mathrm{d} xx^{1 -
		2 \Delta_b} \sin (x).
\end{eqnarray}
For $\frac{1}{2} < \Delta_b < \frac{3}{2}$, the integral converges. Then,
\begin{equation}
	G_B (\vec{q}\,) = \alpha_b \mathcal{P}_b  \sin (\pi \Delta_b) \Gamma (2 - 2 \Delta_b)  q^{2 \Delta_b - 3} . 
\end{equation}
The Fourier transform for $G_F (\vec{q}\,)$ follows from the substitution of $\Delta_b \to \Delta_f + \frac{1}{2}$ and taking derivatives $- i \partial/\partial q_{\mu}$ on $G$. It is useful to note that $\partial_{\mu}q = q_{\mu}/{q}$, and $\partial_{\mu}^2 q^s = sq^{s - 2} [1 + (s - 2) q_{\mu}^2 / q^2]$. This gives near K,
\begin{equation}
	G_{F,\mathrm{K}} (\vec{q}\,)  = i 2 \alpha_f  (1-\Delta_f) \cos (\pi \Delta_f)  \Gamma(1 - 2 \Delta_f) q^{2 \Delta_f - 4}  \cancel{q} .
\end{equation}

We can now check these ansatzes with the self-energy equations. To compute them, we first denote the elementwise product by $\circ$, such that
\begin{equation}
	M = A \circ B \equiv M_{i j} = A_{i j} B_{i j.}
\end{equation}
This is also known as the Hadamard product. Then,
\begin{eqnarray}
	\Sigma_B (\vec{r}\,) & = & [g_C (- \vec{r}\,)]^{\,\scriptscriptstyle{T}}  \! \circ G_F (\vec{r}\,) = - \frac{1}{16 \pi^2} \frac{ \alpha_f}{r^{2 \Delta_f + 4}} 2 \left(
	\cancel{r} \circ \bcancel{r} \right) \nonumber\\
	& = & - \frac{1}{8 \pi^2} \frac{ \alpha_f}{r^{2 \Delta_f + 4}} 
	\begin{pmatrix}
		\tau^2 & x^2 + y^2\\
		x^2 + y^2 & \tau^2
	\end{pmatrix}.
\end{eqnarray}
The Fourier transform for $\Sigma_B$ is,
\begin{eqnarray}
	\Sigma_B (\vec{q}\,) & = &  \frac{\alpha_f}{2\pi}  (2 \Delta_f + 1) \Gamma (- 2 - 2 \Delta_f) \sin (\pi \Delta_f) q^{2 \Delta_f - 1}  \nonumber\\
	&  & \hspace{-4em} \times \begin{pmatrix}
		1 + \frac{\omega_m^2}{q^2}  (2 \Delta_f - 1) & 2 + \frac{q_x^2 + q_y^2}{q^2}  (2 \Delta_f - 1)\\
		2 + \frac{q_x^2 + q_y^2}{q^2}  (2 \Delta_f - 1) & 1 + \frac{\omega_m^2}{q^2}  (2 \Delta_f - 1)
	\end{pmatrix} \!. 
\end{eqnarray}
To satisfy $G_B\Sigma_B=-\mathbb{1}$ at the low energy fixed point, one only need $\mathcal{P}_b G_B \Sigma_B \mathcal{P}_b=\mathcal{P}_b$. This is satisfied with
\begin{equation}
	\mathcal{P}_b \Sigma_B (\vec{q}\,) \mathcal{P}_b =  \frac{\alpha_f}{4\pi} \sin (\pi \Delta_f) \Gamma (-2 \Delta_f)  q^{2 \Delta_f - 1} \mathcal{P}_b .
\end{equation}
For $\Sigma_F$, we first note that $\mathcal{P}_b \circ \cancel{r}=U \cancel{r} U^{\dagger}/2$, and similarly for $\bcancel{r}$. Then, showing $U$ explicitly,
\begin{eqnarray}
	\Sigma_F (\vec{r}\,) & = & - \gamma g_C (\vec{r}\,) \circ G_B (\vec{r}\,) \\
	& = & \frac{- \gamma \alpha_b}{32 \pi^2 r^{2 \Delta_b + 3}}  \Big( e^{- i \mathbf{K}\mathbf{r}} U \cancel{r} U^{\dag} + e^{i \mathbf{K} \mathbf{r}} U \bcancel{r} U^{\dag} \Big)  .\nonumber
\end{eqnarray}
We can drop the $U$'s with our choice $\vartheta=0$. The Fourier transform then gives that near $\mathrm{K}$,
\begin{equation}
	\Sigma_{F,\mathrm{K}} (\vec{q}\,) = -  i  \frac{\alpha_b}{4\pi} \gamma \Delta_b \cos (\pi \Delta_b) \Gamma (- 1 - 2 \Delta_b)  q^{2 \Delta_b - 2} \cancel{q} . 
\end{equation}
Thus, we conclude that our conformal ansatz is consistent with the large-$N$ equations. Similar variation of the solutions are seen in numerics for different $\gamma$'s, and the conformal ansatzes for different exponents under the constraint $\Delta_f+\Delta_b=2$. Deriving exact values of the exponents is challenging due to divergent integrals involved to switch between real and reciprocal space-time for the self-energy and Dyson equations at all $\gamma$. It is left for future work.

\subsection{Scaling behaviors \label{ss:app:scaling}}

\begin{figure}
	\includegraphics[width=\linewidth]{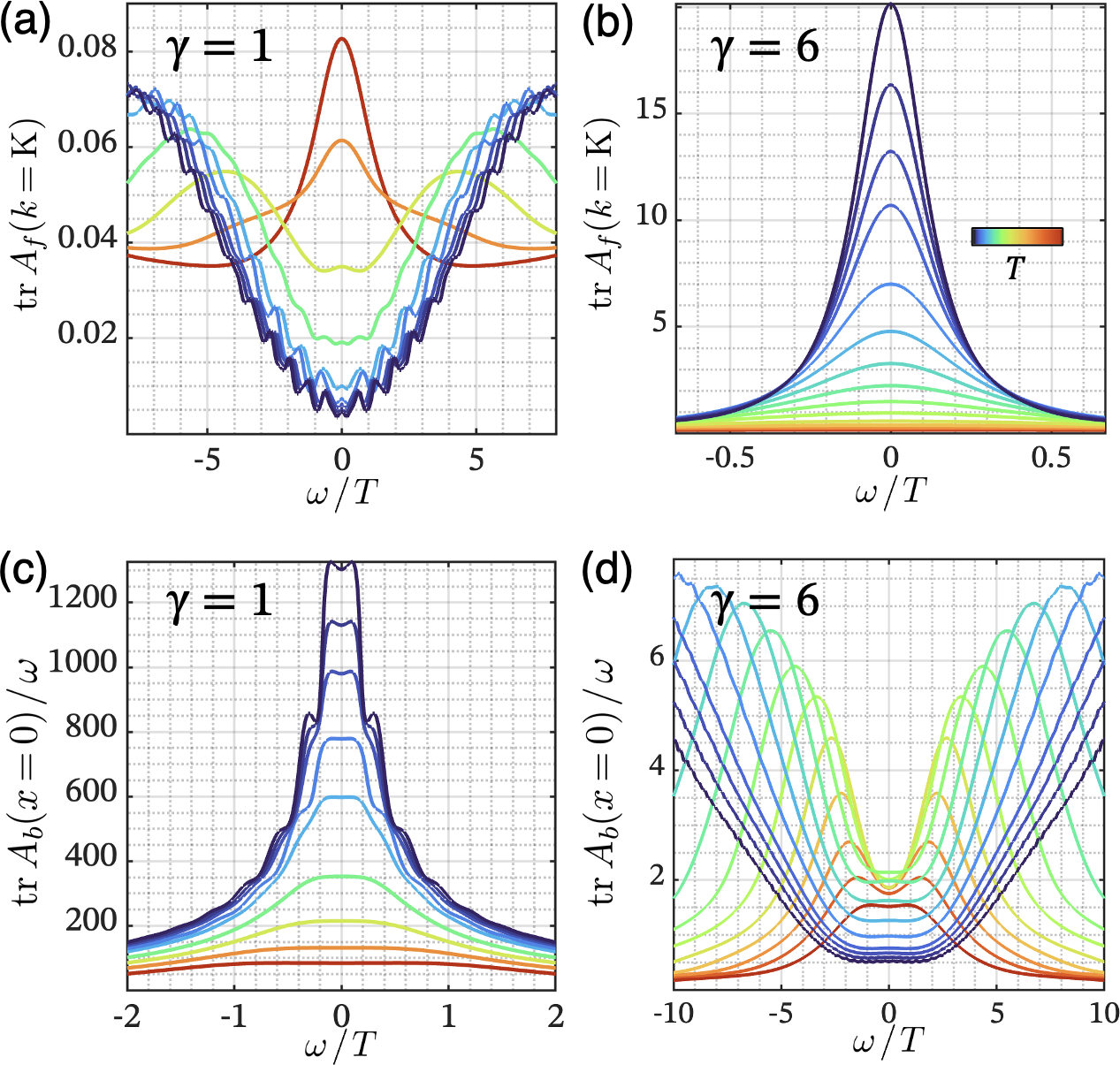}
	\caption{\label{fig:app:scaling} Temperature-scaling behavior of the spectral functions of (a--b) $\mathrm{tr} A_f(k \!\! = \!\! \mathrm{K},\omega)$, and (c--d) $\mathrm{tr} A_b(x \!\! = \!\! 0,\omega)/\omega$, for the gapless phases at different $\gamma$'s. Note that $T$-scaling exponent changes sign across $\gamma$. The ripples at low temperature are finite-size effects. For both cases $J_K/t_c=6$, and $t_f/t_c=-0.2$.}
\end{figure}

Our low-temperature numerical solutions confirm our conformal ansatzes in Eqs.\,(\ref{eq:Gb}). Near the fixed point, the form of the 2D Green's function is governed by the scaling hypothesis
$G(\mathbf{r},\tau;T) \sim T^{2\Delta}h(T \mathbf{r},T \tau)$. In the presence of Lorentz symmetry, it becomes $G(\vec{r};T) \sim T^{2\Delta}\breve{h}(T \vec{r}\,)$. It follows that the scaling of spectral functions as shown in Fig.\,\ref{fig:app:scaling} reveals $\Delta_{b,f}$. Although conformal ansatzes would imply that $h\sim\breve{h}$ is a power law, in a UV-complete theory, the behavior at $\|\vec{q}\,\|=0$ is constrained by sum rules, e.g., $ \int A_f \, \mathrm{d}\omega = 2\pi$ on each sublattice. Consequently, while scaling ansatzes contain power law divergences, they are absent in the UV-complete numerical solutions. Indeed, inspecting Eq.\,\eqref{eq:app:ASb} reveals that if $A_f$ is not quickly diverging, $A_b\vert_{\omega\sim0} \sim \omega$. As a result, the bosonic holons $A_b(\omega)_{nn}$, odd due to p-h symmetry in our studies, is always an extremum when divided by $\omega$ regardless of $\Delta_b$. Thus, we use $T$-scaling behaviors at $\|\vec{q}\,\|\sim0$ to extract the scaling exponents plotted in Fig.\,\ref{fig2}(d). That $T$-scaling changes with $\gamma$ from diverging to vanishing makes for a clear anchor for the exponents.

\begin{figure}
	\includegraphics[width=\linewidth]{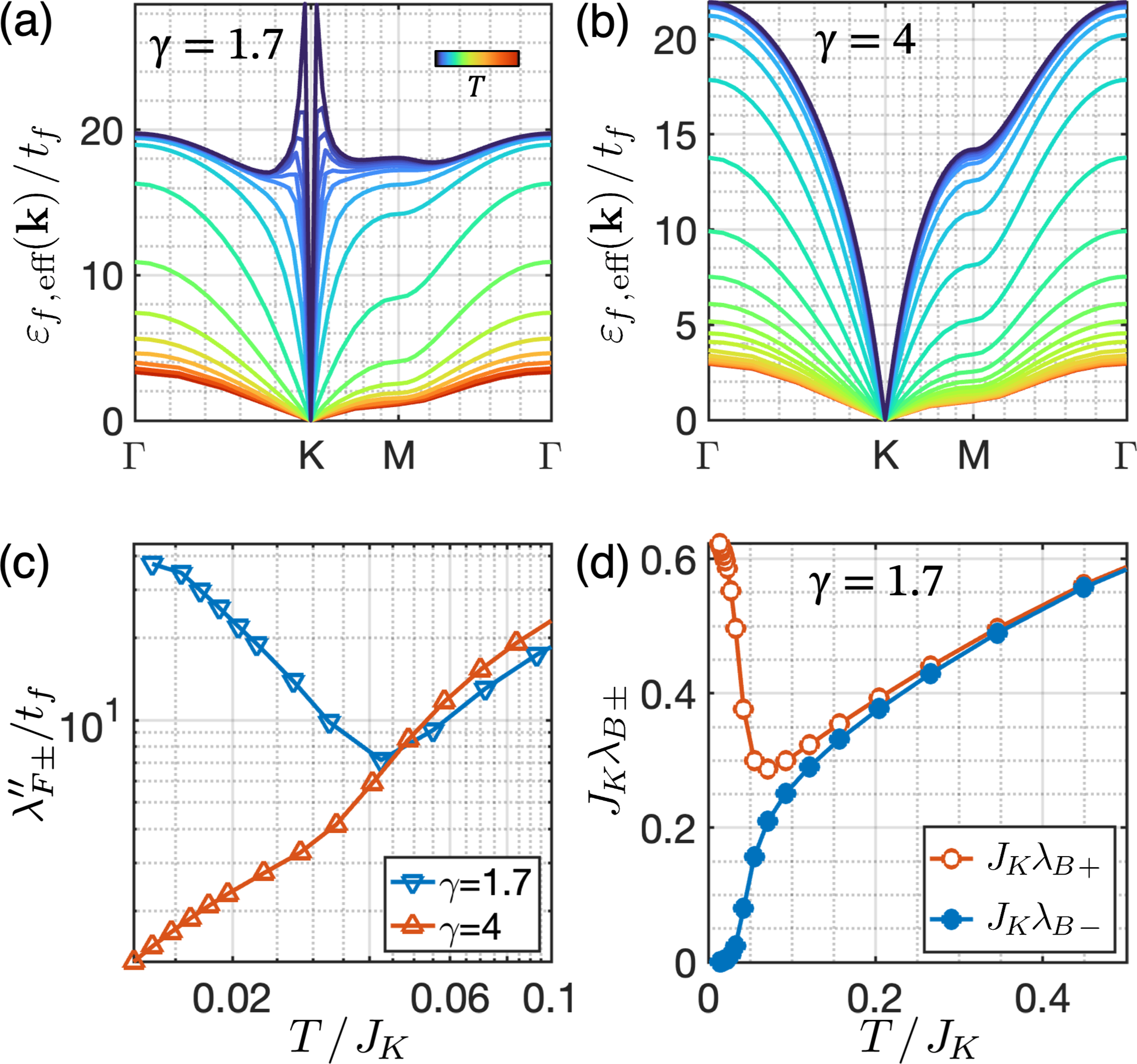}
	\caption{\label{fig:app:Eeff} (a--b) Spinon effective energies $\varepsilon_f$ extracted from real parts of the eigenvalues $\lambda_{F\pm}$of $-G_F^{-1}(k,0+i\eta)$ at (a) $\gamma=1.7$ and (b) $\gamma=4$, as the temperature cools down. Only positive branches are shown. (c) The imaginary part of the eigenvalues, $\lambda^{\prime\prime}_{F{\pm}}$ vs $T$. It is growing for $\gamma=1.7$ where $\Delta_f>3/2$, and vanishing for $\gamma=4$ where $\Delta_f<3/2$. (d) The eigenvalues $\lambda_{B{\pm}}$ at $\gamma=1.7$, which are both nonnegative. The lesser one gives $J_{K,\mathrm{eff}}^{-1}$ while the greater one remains finite at $T=0$. For all data $J_K/t_c=6$, and $t_f/t_c=-0.2$.}
\end{figure}

Another important scaling behavior of the Green's functions lies in their effective energies.
The low temperature fixed point in the overscreened Kondo system is marked by the cancellation of bare energies, $\varepsilon_f$ and $-1/J_K$ by parts of the self-energies $\Sigma_F$ and $\Sigma_B$ respectively, at the critical momenta at $\omega \! = \! 0$~\cite{Ge22,Parcollet1997}. The remaining self-energies, $\delta\Sigma$'s, constitute the conformally invariant Green's functions, $-G^{-1}  =  \delta\Sigma$. The honeycomb lattice has two orbitals per cell. This makes room for nontrivial cancellation of the bare energies as well as the remaining effective energies, $\varepsilon_{f,\mathrm{eff}}$ for spinons and $1/J_{K,\mathrm{eff}}$ for holons. They are given by the real parts of the eigenvalues $\lambda_{\pm}$ of $-G^{-1}$.

In the case of spinons, $\lambda_{F\pm}$ have opposite real parts and identical imaginary parts due to p-h symmetry, and $\varepsilon_f= \lambda^{\prime}_\pm$. The magnitudes $\left|\lambda_\pm\right|$ can be diverging or vanishing governed by the sign of $3-2\Delta_f$, cf.\ Fig.\,\ref{fig2}(d). However, $ \varepsilon_f(\mathbf{k})$ at K and $\mathrm{K}'$ is always zero, as seen in Figs.\,\ref{fig:app:Eeff}(a) and \ref{fig:app:Eeff}(b). When $\Delta_f>3/2$, this leads to an interesting $\mathbf{k}$-dependence of $\varepsilon_f$. As $\mathbf{k}$ approaches $\mathrm{K}$ or $\mathrm{K}'$, $\varepsilon_f$ first tend to diverge but plunges to zero at the critical momenta. Meanwhile, the imaginary parts $\lambda^{\prime\prime}_\pm$ becomes diverging, as shown in Fig.\,\ref{fig:app:Eeff}(c). The same inequality for exponents also affects whether the spectral function is growing or vanishing as $\omega\to0$ at the critical momenta (Fig.\,\ref{fig:app:scaling}). Finally, we note that similar to our previous study in 1+1D \cite{Ge22}, Figs.\,\ref{fig:app:Eeff}(a) and \ref{fig:app:Eeff}(b) show that spontaneous spinon dispersion and dispersion amplification at lower $T$ also occur in 2+1D.

In the case of holons, $\lambda_{B\pm} > 0$ in our studies due to Bose statistics, which requires that $\mathrm{sgn} [A_B(\omega)]_{nn}=\mathrm{sgn}(\omega)$. However, under p-h symmetry energy of the bosons must come in positive-negative pairs. In fact, $\lambda_{B\pm}$ for bosons only give the absolute values of energies, as demonstrated by a simple example:
\begin{eqnarray}
	G(\omega+i\eta) &=&\frac{1}{(\omega+i\eta)^2-\lambda^2} \nonumber\\
	&=& \frac{1}{2m}\left(\frac{1}{\omega+i\eta-\lambda}-\frac{1}{\omega+i\eta+\lambda}\right) \nonumber\\
	&=& \mathrm{P}\frac1{\omega^2-\lambda^2} -i \pi \frac{|\lambda|}{\omega} \delta(\omega^2-\lambda^2).
\end{eqnarray}
Thus, the spectral weight peaks at $\pm\lambda_{B+}$ and $\pm\lambda_{B-}$.
Numerical solutions show that only the lesser eigenvalue $\lambda_{B-}$ vanishes at $T=0$, as shown in Fig.\,\ref{fig:app:Eeff}(d). Its eigenvector is the sublattice-bonding state $(\begin{matrix} 1 & 1\end{matrix})^{\scriptscriptstyle \mathrm{T}}/\sqrt{2}$ in the sublattice basis. Therefore, only sublattice-symmetric Kondo screening takes effect at the fixed point, while the other sector is gapped.

\section{Overscreened Kondo phase diagram with gapped conduction channels \label{ss:app:gapped}}

In this section we present additional numerical results and discussions for the overscreened Kondo honeycomb lattice with Haldane mass-gapped conduction channels (Sec.\,\ref{ss:gapped} in the main text).

\subsection{Free holon population and holon gap \label{ss:app:bgap}}

\begin{figure}
	\includegraphics[width=\linewidth]{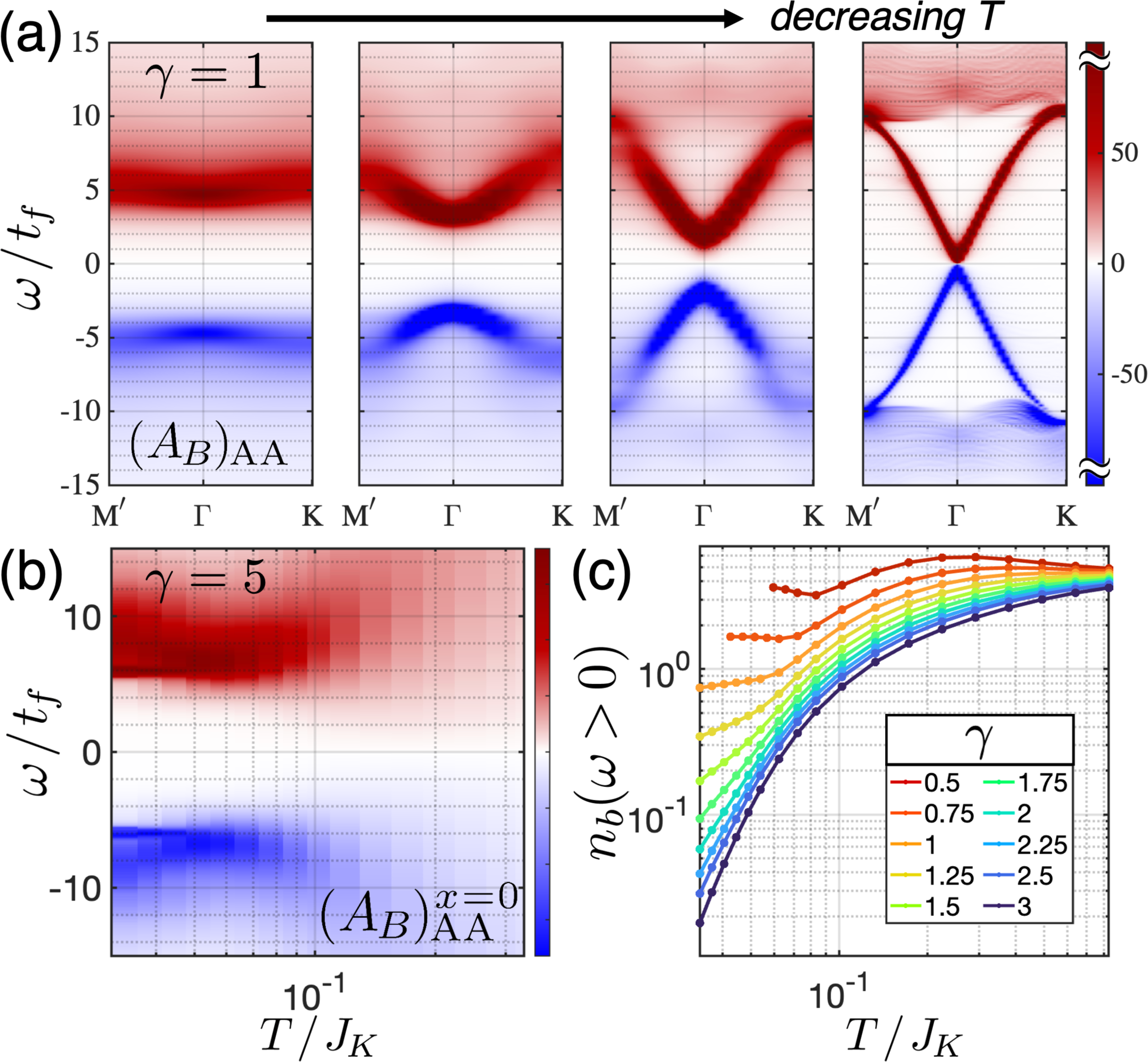}
	\caption{\label{fig:app:holongap}(a) Temperature evolution of the spectral function of holons in Phase I at $\gamma=1$, at $T/J_K=0.083,0.062,0.054$, and 0.037. (b) Temperature evolution of the local spectral function of holons in Phase II at $\gamma=5$, corresponding to Fig.\,\ref{fig3} in the main text, from which the same parameters are used: $\gamma=5$, $J_K/t_c=6$, $|t_c'/t_c|=0.5$, and $t_f/t_c=-0.2$. (c) Population of free holons across different $\gamma$'s, corresponding to Fig.\,\ref{fig4}(b) in the main text, from which the same parameters are used: $J_K/t_c=6$, $|t_c'/t_c|=0.5$, and $t_f/t_c=-0.4$.}
\end{figure}

Here we show the temperature evolution of holon spectral function and the free holon population to supplement the discussion in Figs.\,\ref{fig3} and \ref{fig4}(b). Figure \ref{fig:app:holongap} shows that as the temperature cools down, a coherent bound state of holons emerges from the continuum, migrates to lower energy and becomes gapless at $\Gamma$. In panel (b) of Fig.\,\ref{fig:app:holongap}, we show that the local holon spectral function in Phase II remains gapped at low $T$, corresponding to Fig.\,\ref{fig3} in the main text. Figure \ref{fig:app:holongap}(c) shows the population of free holons across different $\gamma$'s, corresponding to Fig.\,\ref{fig4}(c) in the main text. It shows again that at low $T$, $n_b(\omega\!>\!0)$ is constant in Phase Ia, vanishes by $T^{\xi}$ in Phase Ib, and depletes by $\exp(-m_b/T)$ due to the holon gap $m_b$ in Phase II.

\subsection{Mean-field hybridization model for Phase~I \label{ss:app:hyb}}

\begin{figure}
	\includegraphics[width=\linewidth]{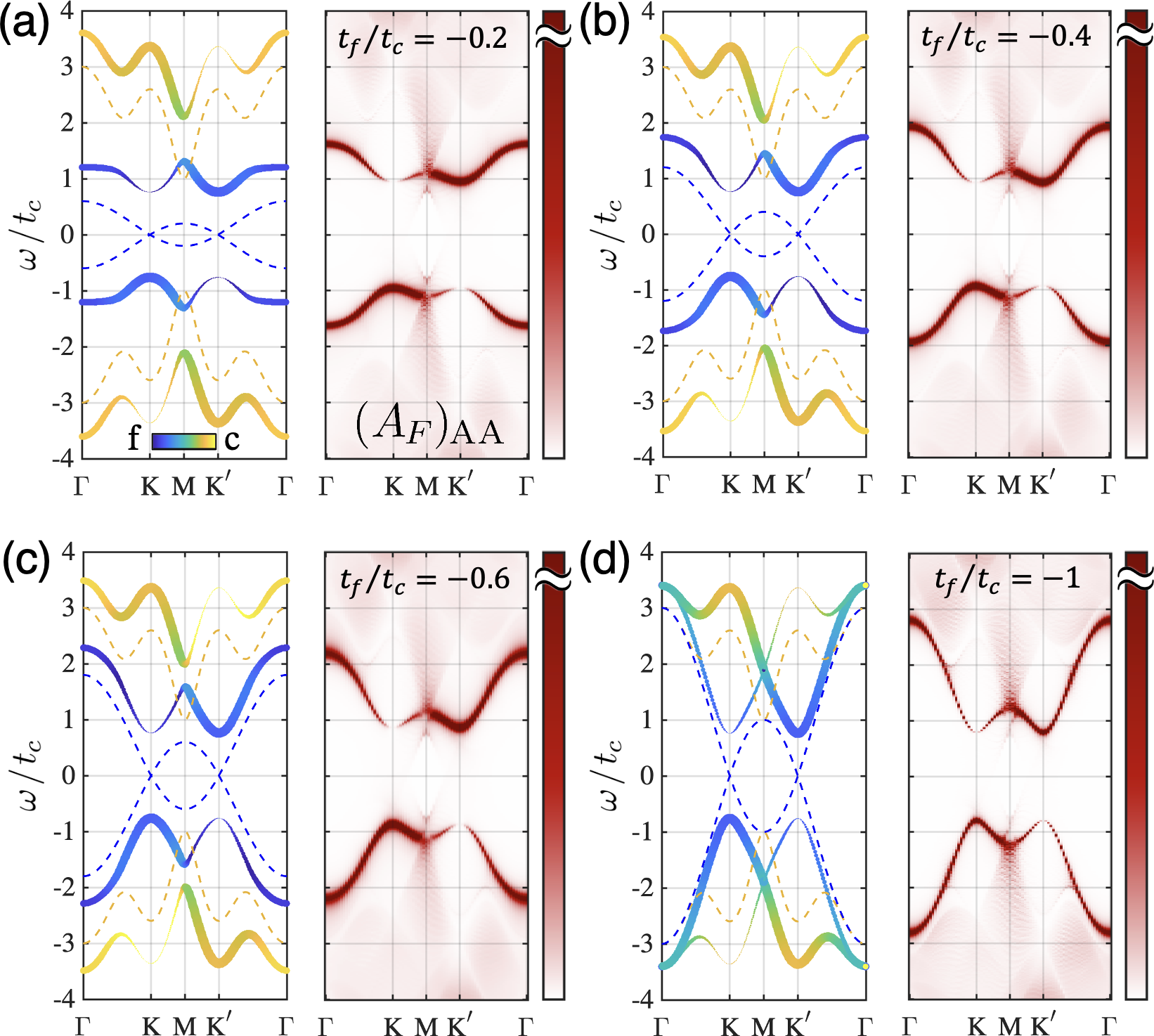}
	\caption{\label{fig:app:hybrid} The energy bands at A sites of the mean-field hybridization model, together with spectral functions $(A_F)_{\mathrm{AA}}$ from dynamical large-$N$ computations, in Phase Ia at $\gamma=0.75$. Data points for the mean-field model energy bands are colored according to their $c$ vs $f$-electron (spinon) contents. Their sizes scale with their weights on A sites. Blue data points trace out the spinon spectral function, which can be compared with $(A_F)_{\mathrm{AA}}$. The bare spinon hopping strengths, $-t_f/ t_c$, are (a) 0.2, (b) 0.4, (c) 0.6, and (d) 1. The hybridization amplitude used is $V/t_c=1.6$, uniform on all sites. Dashed lines are the bare dispersions of $c$- and $f$-electrons. For all cases, $|t_c'/t_c|=0.5$, $J_K/t_c=6$, and $T/J_K=0.04$.}
\end{figure}

In Phase I, spinons are gapped due to dynamic mass generation and the hybridization between $c$- and $f$-electrons, whilst holons develop a coherent gapless mode at $k=\Gamma$. The hybridization can be modeled by a uniform coupling $V c\dg_i f\dn_i+\mathrm{h.c}$. As discussed in the main text, this is the only relevant term that may gap out $f$-electrons. We compare the energy bands of such a mean-field hybridization model with the dynamical large-$N$ spectral functions in Fig.\,\ref{fig:app:hybrid} at various $t_f$, with a hybridization $V/t_c=1.6$. One can see the similarity between the resulting spectral functions. The value of $V$ is chosen to match the frequency centers of spinon spectral bands.

Note that the hybridization strongly modifies the $f$-electron dispersion at M. This is absent in for the Haldane mass gap in Phase II. In the Haldane model, contributions from a purely imaginary NNN hopping vanishes along $\overline{\Gamma\mathrm{M}}$, and $\epsilon_{\Gamma}=3\epsilon_\mathrm{M}$ always. Hence, one can easily discern a Haldane mass gap from a hybridization gap.

\begin{figure}
	\includegraphics[width=0.52\linewidth]{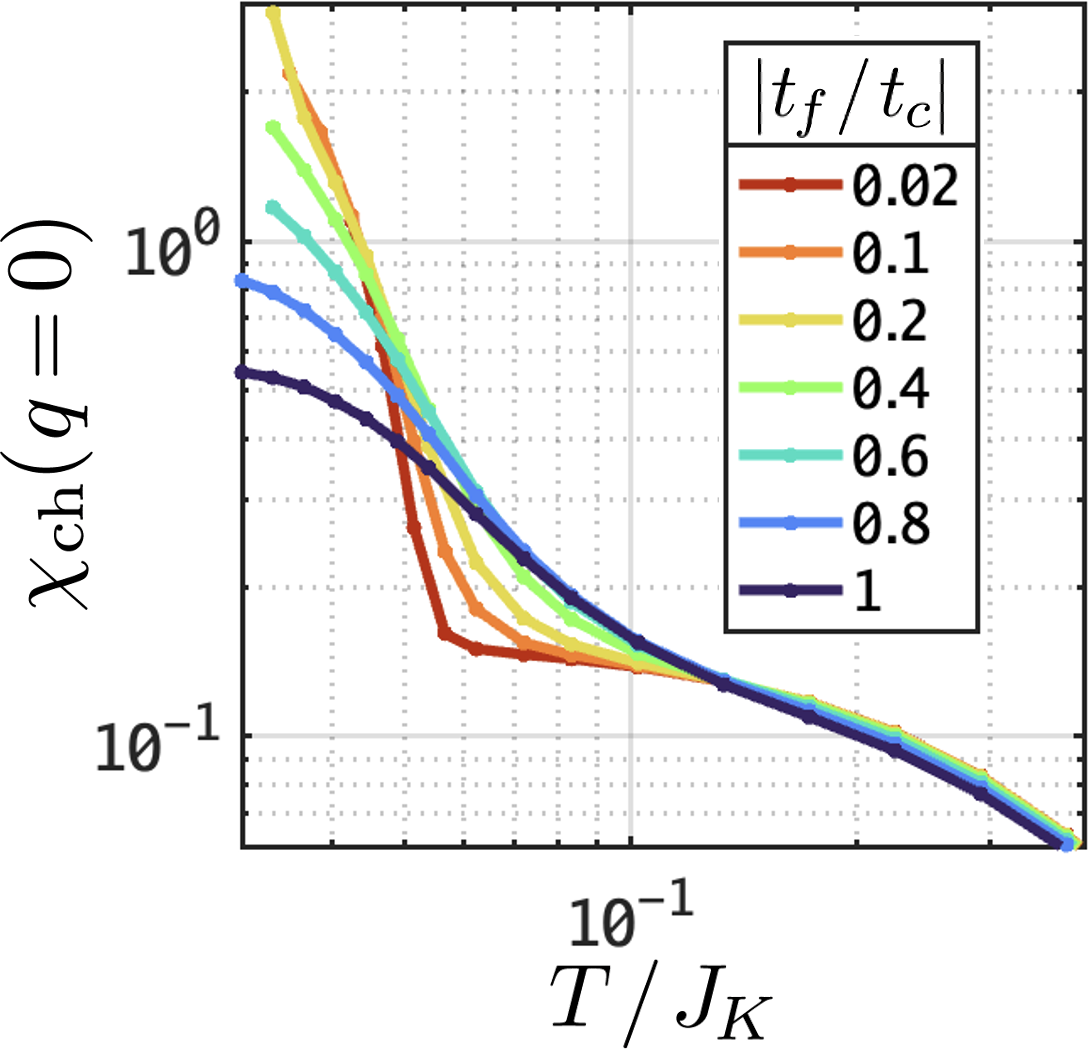}
	\caption{\label{fig:app:tfphase}Channel susceptibility across different $t_f$'s at a constant $\gamma=1.25$, and with $J_K/t_c=6$.}
\end{figure}

\begin{figure}
	\includegraphics[width=\linewidth]{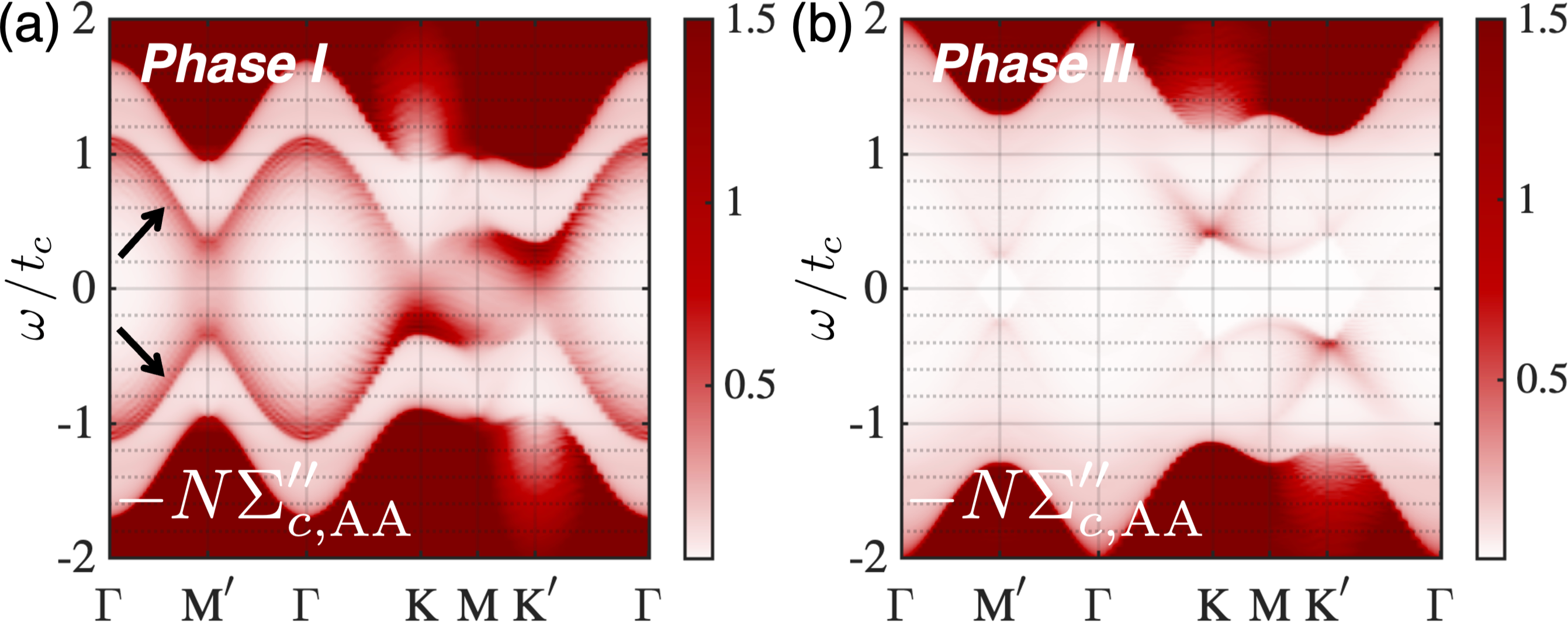}
	\caption{\label{fig:app:Sigc}Conduction-electron self-energy $-(N\Sigma_{c})_\mathrm{AA}^{\prime\prime}$ in (a) Phase I, $\gamma=1.5$, and (b) Phase II, $\gamma=3$. In Phase I the self-energy has two bands of poles dispersing throughout the BZ, indicated by black arrows in (a). The parameters used are the same as in Fig.\,\ref{fig4}(c)(d), $J_K/t_c=6$, $t_f/t_c=-0.2$, and $|t_c'/t_c|=0.5$. The temperatures are at $T/J_K=0.03$.}
\end{figure}

\subsection{Phase diagram at constant $\gamma$ and varying $t_f$ \label{ss:app:tf}}

The uniform mean-field value for the bond variables $t_f$ determines the strength of antiferromagnetic Heisenberg coupling $J_H$ in the Abrikosov fermion representation of spins, which can be derived from Eq.\,\eqref{eq3} using the saddle-point condition ${\delta}F/{\delta}t_f^*=0$. Results in the main text are at constant $t_f$'s, which at low $T$ is equivalent to a constant $J_H$, and a constant $T_K/J_H$ when $J_K$ is constant. (Alternatively, one can fix $J_H$ and update $t_f$ at each $T$ \cite{Komijani18}.) In the TR invariant regime, i.e.\ with only NN hopping $t_c$, tuning $t_f$ does not change our conformal ansatzes nor the scaling exponents. Its only effect is to move the system away from the local Kondo impurity fixed point at high temperature~\cite{Ge22,Parcollet1997}. A large $t_f$ will keep the spinons dispersive at all $T$ and eliminate any trace of this local fixed point where both spinon and holon Green's functions become essentially local.

In the TR broken regime, i.e.\ when both $t_c,t_c'$ are nonzero, tuning $t_f$ has a similar effect to increasing $\gamma$ in the phase diagram, as seen in Fig.\,\ref{fig:app:tfphase}. Although smaller $t_f$ delays the departure from the local Kondo fixed point at higher $T$ when the temperature cools down, close to zero temperature we see that an increasing $t_f$ changes the uniform channel susceptibility from divergent to regular. The magnetic susceptibility is always regular at low $T$ due to the spinon gap.

\subsection{Electron self-energy across the phase diagram \label{ss:app:sigc}}

Another distinct feature across the phase diagram is the self-energy of conduction electrons, $N\Sigma_c$.  As discussed in the main text, only for Phase I where $\gamma<\gamma_c\sim2$ the $c$-$f$ hybridization is active. This results in a resonance in the conduction-electron self-energy $\Sigma_c$. As was shown in Figs.\,\ref{fig4}(c) and \ref{fig4}(d), at K points, there are poles in $N\Sigma_c$ crossing zero frequency as the $T$ goes down in Phase I, which is absent in Phase II. Momentum cuts of $N\Sigma_c(k,\omega+i\eta)$ in Fig.\,\ref{fig:app:Sigc} further demonstrate the difference.

In Phase I, $N\Sigma_c$ has two sharp ``bands'' of poles throughout the BZ at the lowest temperatures. Thus, the self-energy in Phase I can be modeled by
\begin{equation}
	N\Sigma_c(k,\mathrm{z}) \sim \frac{V^2}{\mathrm{z}\mathbb{1}-H(k)},
\end{equation}
where $H(k)$ describes the dispersion of these bands, and $V$ is the hybridization amplitude between $c$ and $f$. The leading terms in $H(k)$, extracted from $N\Sigma_c(k,0+i\eta)$, also resembles a Haldane model with a purely imaginary NNN hopping. Such bands are absent in Phase II, shown in Fig.\,\ref{fig:app:Sigc}(b). In addition, we find that there is a ``band inversion'' near K points in Phase I as the system cools down from high temperature. This corresponds to the zero-crossing of the pole at K in Fig.\,\ref{fig4}(c).

Finally, these divergent poles in Phase I can survive the large-$N$ suppression even though nominally $\Sigma_c \sim \mathrm{O}(1/N)$. On the other hand, $\Sigma_c$ in Phase II is incoherent throughout most of the BZ.

\bibliography{KL}

\end{document}